\documentclass[12pt,preprint,apj]{emulateapj}
\usepackage{epstopdf}
\usepackage{color}

\newcommand{\kms}{{km s$^{-1}$}}

\shorttitle{M105 stars}
\shortauthors{Lee \& Jang}

\begin{document}

\title{
%{\color{red} {\bf Revised  \today}} \\
Dual Stellar Halos in the Standard Elliptical Galaxy M105 and Formation of Massive Early-Type Galaxies  }
 
\author{Myung Gyoon Lee \&  In Sung Jang}
\affil{Astronomy Program, Department of Physics and Astronomy, Seoul National University, 1 Gwanak-ro, Gwanak-gu, Seoul 151-742, Korea}
\email{mglee@astro.snu.ac.kr, isjang@astro.snu.ac.kr }

%==============================================================================================================

\begin{abstract}
M105 is a standard elliptical galaxy, located in the Leo I Group. 
We present photometry of the resolved stars 
in its inner region at $R \approx 4\arcmin \approx 4R_{\rm eff}$, 
obtained from  F606W and F814W images  in the {\it Hubble Space Telescope}  archive. 
We combine this with photometry of the outer region at $R\approx 12\arcmin\approx 12R_{\rm eff}$ from archival imaging data.
Color-magnitude diagrams of the resolved stars in the inner region show a prominent red giant branch (RGB) with a large color range, while those of the outer region show better a narrow blue RGB.
The metallicity distribution function (MDF) of the RGB stars shows the existence of two distinct subpopulations: a dominant metal-rich population (with a peak at [M/H]$\approx0.0$) and a much weaker metal-poor population  (with a peak at [M/H]$\approx-1.1$). 
The  radial number density profiles of the metal-rich and metal-poor RGB stars are fit well by a S{\'e}rsic law with $n=2.75\pm0.10$  and $n=6.89\pm0.94$, 
and by a single power law
($\sigma \propto R^{-3.8}$ and $\sigma \propto R^{-2.6}$), respectively. 
The MDFs of the inner and outer regions can be described well by accretion gas models of chemical evolution with two components.
These provide strong evidence that there are two distinct stellar halos in this galaxy, metal-poor and red metal-rich halos, consistent with the results based on globular cluster systems in bright early-type galaxies (Park \& Lee 2013).
We discuss the implications of these results with regard to
the formation of massive early-type galaxies in the dual halo mode formation scenario.

\end{abstract}

%, one magnitude deeper than the previous study. % studied before. 
%$n=1.9\pm0.3$  and $n=2.8\pm0.8$ 
%(similar to the de Vaucouleurs law), respectively. The latter is also fit well 

\keywords{galaxies: distances and redshifts --- galaxies: individual (M105) --- galaxies: stellar content---galaxies:structure --- galaxies:elliptical and lenticular, cD --- galaxies:halo}

\section{Introduction}

Globular clusters are an excellent probe to investigate the halos in nearby galaxies.
Color distributions of globular clusters in early-type galaxies (ETGs), including elliptical and lenticular galaxies,
show that there are two subpopulations: blue (metal-poor) and red (metal-rich) globular clusters \citep{gei96,bro06,pen08}. 
Recently \citet{par13} investigated the spatial distribution of these subpopulations in 23 Virgo ETGs using a homogeneous data set based on the ACSVCS catalog, which was derived from the observations with the Hubble Space Telescope (HST) \citep{jor09}. They found that the ellipticity of
the red globular cluster systems (GCSs) follows much more tightly than that of the stellar light, compared with the ellipticity of the blue GCSs. 
From this and other related properties of the GCSs they suggested that massive ETGs may have dual halos: a blue (metal-poor) halo
and a red (metal-rich) halo. The red halos are much more elongated and centrally concentrated than the blue halos. The red halos may have stronger rotation than the blue halos. They predicted that ETGs may host stellar halos corresponding to these two halos. However, the number density of the blue halo stars in ETGs is much lower than that of the red halo stars in the outer region of their host galaxies so that it is not easy to see the blue halo stars in typical images of ETGs. Typical images of ETGs show only the structure of the red halos that dominate the integrated light. 

Indeed the existence of blue (metal-poor) RGB stars in ETGs was studied for a few nearby galaxies: NGC 3115 (S0), NGC 5128 (Ep), NGC 3377 (E5), and M105 (NGC 3379, E1) \citep{els97,har02,rej05, rej11,crn13,har07a,har07b,bir15,pea15}. 
In her pioneering study of stars in an $HST$/WFPC2 field $5\arcmin$ (22 kpc) away from the major axis of NGC 3115, an edge-on bright S0 galaxy, \citet{els97} found that the $(V-I)$ color distribution
of the resolved red giants in this field shows a clear bimodality. Then she pointed out that there are two distinct halo
populations of roughly equal size: a metal-poor halo with [Fe/H] $\approx -1.3$ and a metal-rich one with [Fe/H] $\approx -0.0$. This bimodality was confirmed later in a recent study of the resolved stars in the more remote fields of this galaxy based on deeper photometry by \citet{pea15}.

%%%%%%%%%%%%%%%%%%%%%%%%%%%%%%%
% Table 1
%%%%%%%%%%%%%%%%%%%%%%%%%%%%%%%
\begin{deluxetable*}{lcl}
\tablewidth{0pc} %\tablenum{2}
\tablecaption{Basic Parameters of M105\label{tab-info}}
\tablehead{
\colhead{Parameter} 	& \colhead{Value} & \colhead{References}}
\startdata
R.A.(J2000), Dec(J2000) 	& 10 47 49.6,  +12 34 54	& 1 \\
Morphological Type 					& E1 							& 2\\
Total magnitude and color  	& $B^T = 10.24\pm0.03$, $( B^T - V^T ) = 0.96\pm0.01$ & 2\\
Ellipticity	 			& 0.11 & 3\\
%Inclination angle 		& ? deg & \citet{bus96}\\
Position angle 			& 71 deg & 3\\
%Position angle 			& 53 deg	& RC3  \\
$D_{25}(B)$				&  $262\arcsec$ & 2\\ 
%$322\arcsec.20 \times 286\arcsec.76$
%$D_{25}(B)$				& $329\arcsec.70$(major), $151\arcsec.66$(minor) & RC3\\
%$D_{25}(K)$				& $429\arcsec.00$(major), $218\arcsec.79$(minor) & Jarrett et al. (2003) \\
Effective radius		& $58\farcs7$ & 2 \\
%Effective radii 		& $24\arcsec.6$, $20.5\arcsec.5$, $22\arcsec.5$ & \citet{bus96} \\
Systemic velocity 		& 911 km s$^{-1}$ & 1\\
%mass(nuclear region)  & $7 \times 10^8  M_\odot$ & \\
%Total mass & & \\
Central velocity dispersion ($R_{\rm eff}/8$) & $213 $ km s$^{-1}$ & 4 \\
Foreground reddening 	& %$A_B=0.089$ $A_V=0.067$ $A_I=0.037$, 
$E(B-V)=0.022$ & 5 \\
Distance 				& $(m-M)_0 = 30.05\pm0.02({\rm ran})\pm0.12({\rm sys})$ & 6 \\
Plate scale 				& 49.6 pc arcsec$^{-1}$ & 6 \\
Absolute total magnitudes 		& $M_B^T = -19.89 $, $M_V^T = -20.83 $  & 6\\ 
\enddata
%\tablenotetext{a~}{caption.}
\tablerefs{(1) NED; (2) \citet{dev91}; (3) \citet{mak14}; (4) \citet{cap13}; (5) \citet{sch11}; (6) This study.}
\label{tab_info}
\end{deluxetable*}

Resolved stars in NGC 5218 %, the nearest S0 galaxy, 
were extensively studied by \citet{har02,rej05, rej11, crn13, rej14} and \citet{bir15}.
%  
%{%\color{red}\bf 
Because of the presence of notable dust lane in the middle, whether the morphological type of NGC 5128 is pecular S0 or E has been controversial. In a recent review, \citet{har10} %(PASA, 27. 475) 
concluded that it is an Ep, noting that its global properties are closer to normal Es, except for the dust lane. %}
NGC 5128 covers a very large angular size in the sky so that deep images were obtained only for a part of the entire galaxy. % \citep{rej05,crn13,bir14}. 
\citet{har07a} studied the resolved stars in NGC 3377, a highly elongated E5 galaxy
with intermediate luminosity ($M_V = -19.9$ mag), finding that the mean metallicity of these stars is between those of
massive galaxies and dwarf galaxies.
 
In this study we investigate the resolved stars  in M105,
a nearby E1 galaxy in the Leo I Group that is located at 10.6 Mpc \citep{lee13}.
M105 is located at the distance that is ideal for studying not only its resolved stars but also their radial distribution, if we use high resolution images.
M105 has been known as one of the standard elliptical galaxies of which the surface brightness profile can be described impressively over many orders of magnitudes by the de Vaucouleurs $R^{1/4}$ law \citep{dev79,cap90,wat14}.
 Basic parameters of M105 are listed in {\bf Table \ref{tab_info}}. 
 Foreground reddening toward M105 is known to be very small, $E(B-V)=0.022$ \citep{sch11}. 
Corresponding values are $A_I=0.037$ and $E(V-I)=0.030$. 
Internal reddening for the old red giants in elliptical galaxies must be low so that it is assumed to be zero in this study.

Resolved stars in M105 were studied previously by \citet{sak97,gre04} and \citet{har07b}.
 \citet{sak97} estimated the distance to this galaxy using the tip of the red giant branch (TRGB) method \citep{lee93}, from
 the photometry of the resolved stars in the  
 $HST$/WFPC2 F814W images of a field $6\arcmin$ west of M105 . 
 However, they had only one-band images so that they had no color information of the resolved stars.  
\citet{gre04} presented F110W($J$) and F160W($H$) photometry of the resolved stars in three small fields at $R=3\arcmin$, 
$R=4\farcm5$, and $R=6\arcmin$, based on $HST$ NICMOS images. They found that the resolved stars have a broad range of metallicity from [Fe/H]$\sim-1.5$ to $+0.8$, with a mean of the solar metallicity.
They pointed out that the stellar population in their fields is similar to that of the bulge of the Milky Way Galaxy.
\citet{har07b} observed a remote western field at  $R=12\arcmin$ ($\approx 12 R_{\rm eff}$), using $HST$/ACS F606W and F814W filters.  
%
%{%\color{red} \bf 
They carried out a better TRGB determination than the previous literature, being able to work with two color photometry. %}
They found that the metallicity distribution function (MDF) of the red giant stars in this field is very broad and flat, 
%{%\color{red} \bf 
showing no dominant population. They stated that the shape of this MDF is not like any known MDFs of other galaxies. %}
They suggested also that the resolved stars in this remote region  is composed of two distinct subpopulations, metal-poor and metal-rich populations, noting that the metal-poor stars get more significant than the metal-rich stars with increasing radius in this outer region.

In this study we present photometry of the resolved stars 
in the inner region ($2\arcmin<R\leq6\arcmin$) of M105,
obtained from  F606W and F814W images  in the $HST$
archive. Then we combine this with the photometry of the outer region studied previously by \citet{har07b}, to investigate the stellar populations over a large radial range.
Section 2 derives photometry of the point sources in the images and 
\S3 presents color-magnitude diagrams (CMDs) of the resolved stars in this galaxy,
derives a distance to this galaxy using the TRGB method,
estimates metallicities of bright RGB stars, presents their MDFs,
and shows the radial distributions of the metal-poor and metal-rich RGB stars.
In \S4 %{%\color{red} \bf 
we investigate the MDFs of the RGB stars in M105 with analytic chemical evolution models, %}  
compare the MDFs of M105 with those of other nearby galaxies, and
suggest an updated dual halo mode scenario for the formation of massive ETGs. 
% implications of these results in \S4, and summarize 
Primary results are summarized in the final section.

%%%%%%%%%%%%%%%%%%%%%%%%%%%%%%%%%%%%%%
%%%  Fig 1
%%%%%%%%%%%%%%%%%%%%%%%%%%%%%%%%%%%%%%
\begin{figure*}
\centering
\includegraphics[scale=0.96]{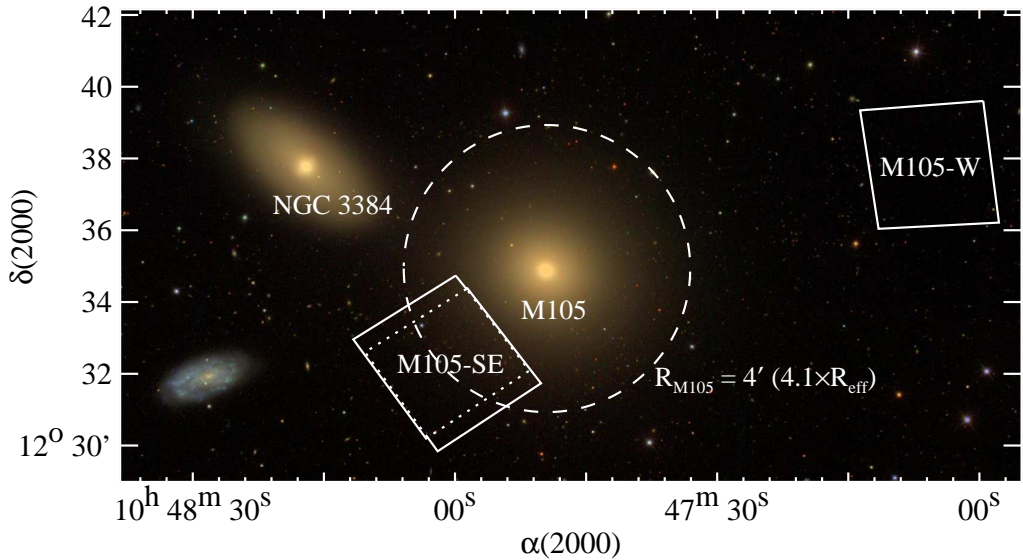} %{finding.eps} %{finding1.eps} %{finding_2.eps} 
\caption{A finding chart for M105 (E1). 
HST fields used in this study are marked on the color map of 
the Sloan Digital Sky Survey: % $i$-band image: 
one inner field M105-SE (at $R\approx 4 R_{\rm eff} \approx 12$ kpc ) and one outer field M105-W  (at $R\approx 12 R_{\rm eff} \approx 36 $ kpc). 
The dashed-line represents a circle with $4\arcmin$ radius centered on M105. 
NGC 3384 (SB0) is also marked in the north east of M105.
A spiral galaxy  in the south-east region, NGC 3389 (Sc(s)), is known to be a background galaxy, being located about 10 Mpc behind M105.} 
\label{fig_finder}
\end{figure*}

%%%%%%%%%%%%%%%%%%%%%%%%%%%%%%%%
% Table 2
%%%%%%%%%%%%%%%%%%%%%%%%%%%%%%%%
\begin{deluxetable*}{lcccrrcc}
%\tabletypesize{\footnotesize} % \footnotesize \scriptsize
%\tabletypesize{\tiny}
\setlength{\tabcolsep}{0.05in}
%\rotate
\tablecaption{A Summary of $HST$ Data for M105} % and NGC 3377 (and the other galaxies?)}
\tablewidth{0pt}

\tablehead{ \colhead{Field} & \colhead{R.A.} &  \colhead{Dec} & \colhead{Instrument} & \multicolumn{2}{c}{Exposure time [sec]} & \colhead{Prop. ID} \\
	& (2000) & (2000)	&	& F606W	& F814W	& &} 
\startdata
M105-SE						& 10 48 01.53	& 12 32 28.2	& ACS/WFC & 9,775 & 9,775 & 10413   \\
							& 10 48 00.20	& 12 32 06.8	& ACS/WFC & 9,775 & 9,775 & 10413   \\
M105-W						& 10 47 05.67	& 12 37 49.8	& ACS/WFC & 38,500 & 22,260 & 9811  \\

\hline

\enddata
%\tablerefs{1. This study. 2. \citet{har07b}. 3. \citet{har07a}.}
\label{tab_data}
\end{deluxetable*}

%%%%%%%%%%%%%%%%%%%%%%%%%%%%%%%%%%%%%%
%%%  Fig 2
%%%%%%%%%%%%%%%%%%%%%%%%%%%%%%%%%%%%%%

\begin{figure}
\centering
\includegraphics[scale=0.6]{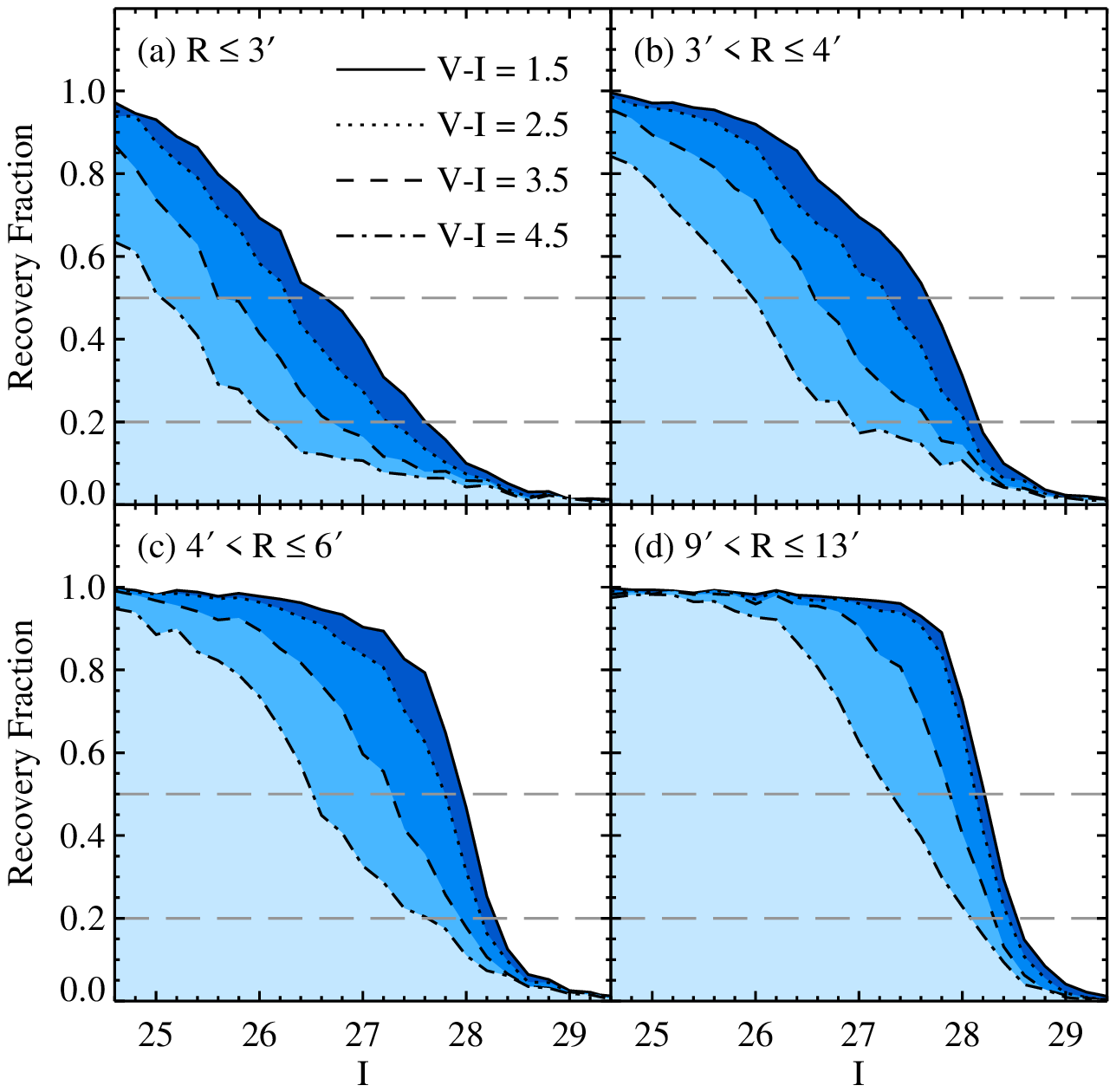} %{comp_m105.eps} %{finding1.eps} %{finding_2.eps} 
\caption{Recovery fractions of artificial stars as a function of $I$-band magnitude: (a) $R\leq3\arcmin$, (b) $3\arcmin<R\leq4\arcmin$, (c) $4\arcmin<R\leq6\arcmin$, and 
(d) $9\arcmin<R\leq13\arcmin$.
Curved lines indicate the recovery rates (completeness) of artificial stars with different $V-I$ color (solid lines for $V-I=1.5$ stars, dotted lines for $V-I=2.5$ stars, short dashed lines for $V-I=3.5$ stars, and dot-dashed lines for $V-I=4.5$ stars, respectively). Upper and lower horizontal dashed lines in each panel represent  50\% and 20\% recovery rates, respectively.
}
\label{fig_comp1}
\end{figure}

\section{Data Reduction and analysis}

\subsection{Data Reduction}

There are three bright galaxies around the center of the Leo I Group:
M105, NGC 3384 (SB0), and NGC 3389 (Sc(s)). 
%{%\color{red}\bf
 NGC 3389 is known to be located about 10 Mpc behind M105 and NGC 3384 both of which are at similar distances \citep{sti09, bos14}. NGC 3389 belongs to another small group in the Leo I Cloud that is behind the Leo I Group \citep{sti09}. %}
%The distance to NGC 3398 is known to be $23.26\pm1.96$ Mpc, based on the expanding photosphere method for SN2009md (Bose & Kumar 2014), so that it is located significantly behind M105.
%Fraser et al. 2011, MNRAS, 417, 1417  
%Bose,S., & Kumar, B. 2014, ApJ, 782, 98 23.26\pm1.96 Mpc EPM for SN2009md
%\bibitem[Fraser et al.(2011)]{fra11}Fraser, M. et al. 2011, \mnras, 417, 1417
%\bibitem[Bose \& Kumar(2014)]{bos14}Bose,S., \& Kumar, B. 2014, \apj, 782, 98                        
%\bibitem[Stierwalt et al.(2009)] 21.4 Mpc, NGC 3389 system is a small grounp in the Leo I Cloud behind the Leo I Ring.
{\bf Figure \ref{fig_finder}} displays a finding chart for  M105, showing the $HST$ fields on the SDSS color image. %$i$-band image.
F606W  and F814W images of two fields in M105
%NGC 3384 taken with the  $HST$/ACS (Proposal IDs: ??) 
are available in the $HST$ archive, as marked in this figure  and listed in {\bf Table \ref{tab_data}} :
 a south-eastern inner field (called the M105--SE field), and a western outer field (called the M105--W field) that was used previously for the study of resolved stars by \citet{har07b}.
The M105--SE field is located at $R\sim 4\arcmin$ from the galaxy center, while the M105--W field is at $R\sim 12\arcmin$.
%We also analyzed the $F606W$  and $F814W$ images of NGC 3377, an E5 galaxy in the Leo I Group. NGC 3377 ($M_V^T=-19.9$ \citep{har07a}) is $\sim 1$ mag fainter than M105 so that it is a useful reference to compare stellar populations in elliptical galaxies with different luminosity (or mass). \citet{har07a} presented photometry of the resolved stars in this galaxy. We derived the photometry of the resolved stars in this galaxy from the same data as used in \citet{har07a}, but using the same procedures as used for the inner region in this study for comparison.
 
We combined the images of the M105--SE and W fields to make a mosaic drizzled image for the entire observed field, using Tweakreg and AstroDrizzle tasks in DrizzlePac provided by the Space Telescope Science Institute (STScI) \\
(http://www.stsci.edu/hst/HST$\_$overview/drizzlepac/), as done in \citet{jan14}.
% We constructed deeper drizzled images for each filter combining the charge transfer efficiency corrected and flat field-calibrated single images (indicated by $\_$flc.fits images)
%flat fielded and charge transfer efficiency corrected single images (indicated by $\_$flc.fits images) 
%We adopted a PIXFRAC value of 0.7 and created final images with a pixel scale, $0\arcsec.03???$ per pixel.
% a PIXFRAC value of 0.7 to generate 
The final images have a pixel scale, $0\farcs03$ per pixel.
Total exposure times are 19,550 s  for each of F606W and F814W. 
These exposure times are slightly shorter than those for the M105-W field, 
but are still long enough to study the resolved stars in the M105-SE field.

%%%%%%%%%%%%%%%%%%%%%%%%%%%%%%%%%%%%%%
%%%  Fig 3
%%%%%%%%%%%%%%%%%%%%%%%%%%%%%%%%%%%%%%
\begin{figure*}
\centering
\includegraphics[scale=0.9]{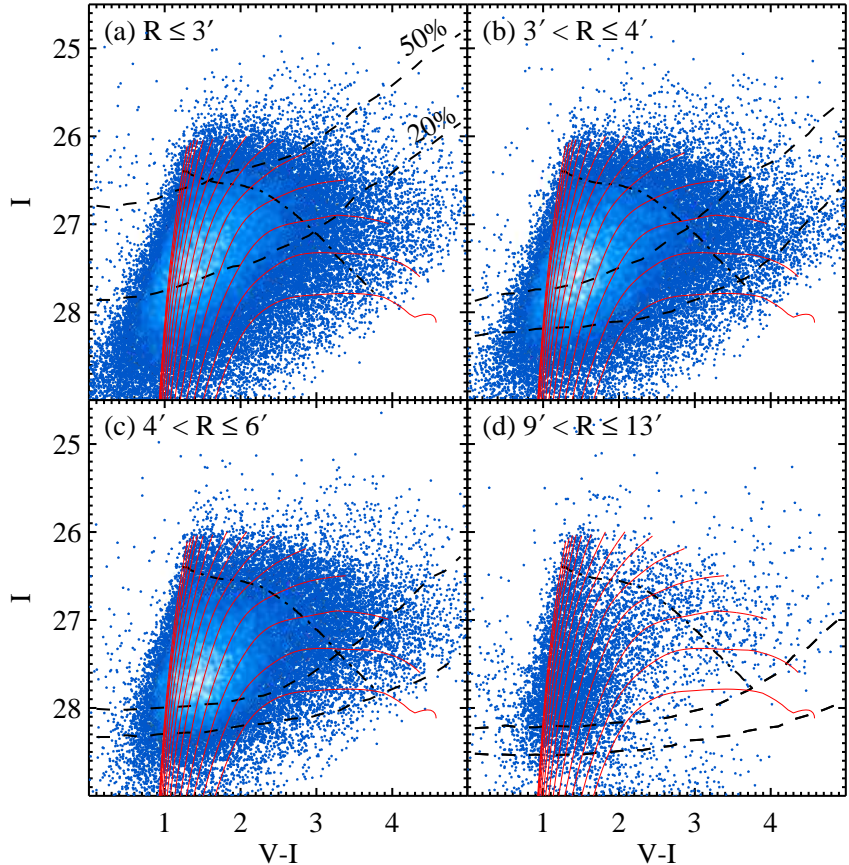} %{cmd_m105.eps} %{fcmd_m105.eps}%%{cmd_m105.eps} %cmd_2.eps} %{testn3384cmd1.eps} 
\caption{$I-(V-I)$ CMDs of the resolved stars in M105:
(a) $R \le 3\arcmin$, (b) $3\arcmin< R  \le 4\arcmin$,
(c) $4\arcmin < R  \le 6\arcmin$, and (d) $9\arcmin<R  \le 13\arcmin$.
Smoothed maps display the number density maps for the high density regions (Hess diagrams).
Red solid lines represent theoretical isochrones for 12 Gyr age, [$\alpha$/Fe]=0.2, and a range of metallicity ([Fe/H] = --2.4 to +0.2 ([M/H] = --2.2 to +0.4 ) with a step of 0.2 from left to right) in the Dartmouth models \citep{dot08}. Upper and lower dashed lines in each panel represent 50\% and 20\% completeness limits, respectively.
} 
\label{fig_cmd}
\end{figure*}

\subsection{Photometry and Completeness Test}

We derived photometry of the point sources in the drizzled images using 
IRAF/DAOPHOT package \citep{ste94}, as done in \citet{jan14}. We selected the point sources using the sharpness parameters provided by DAOPHOT.
Then we converted the instrumental magnitudes of the sources to the standard Johnson-Cousins $VI$ Vega magnitudes, following \citet{sir05}. 

We estimated the completeness of our photometry using artificial star experiments.
First, we selected four $0.5\arcmin \times 0.5\arcmin$ subregions at different galactocentric radii. 
We generated artificial stars with colors $(V-I)=0.1 - 4.9$ (in steps of 0.2 mag) and magnitudes  $I=24.5-30.0$ mag. Then we added $\sim$5000 (about 10 \% of the detected point sources) artificial stars in each sub-region,  
and iterated this procedure four  times to increase the total number of artificial stars.
We estimated the recovery rate (completeness) from the ratio of the number of the recovered stars and that
of the added stars. 
%Total N of artificial stars : $\sim$5000 stars  $\times$  4 (sub-regions) $\times$  4 (iteration) $\times$ 25 (N of color bin) $\times$ 2 (M105, N3377) = $\sim$4 million stars\\
{\bf Figure \ref{fig_comp1}} presents the completeness as a function of $I$ magnitudes for four color values ($(V-I)=1.5, 2.5, 3.5$, and 4.5).
%, and {\bf Figure \ref{fig_comp2}} displays the completeness map in the CMD for visual aid.
The 50\% completeness levels for $(V-I)=3.5$ are $I\approx 28.0$ mag at the outer region, and they get brighter as the color gets redder or the galactocentric distance decreases. 
%We used 2-$\sigma$ as the detection threshold, and derived the PSFs using isolated bright stars in the images.  
%We derived aperture corrections using a large aperture
%with radius of $1\arcsec$ for several isolated bright stars in the images. 
%The uncertainties associated with aperture correction are on average 0.02 mag for both filters.
%We transformed the instrumental magnitudes 
%into the standard Johnson-Cousins $VI$ magnitudes, 
%following equations (1) and (12) for observed magnitudes of \citet{sir05}. 
% The uncertainties associated with the photometric transformations   are, on average, 0.02 mag. 
% 
\section{Results}

\subsection{CMDs of the Resolved Stars}

In {\bf Figure \ref{fig_cmd}} we display %color-magnitude diagrams (
CMDs of the resolved stars in the four ranges of projected galactocentric distance %($R$) 
%inner ($2\arcmin.5<R<3\arcmin.5$) and outer  ($R>3\arcmin.5$) regions 
of M105: (a) $2\arcmin < R  \le 3\arcmin$, (b) $3\arcmin < R  \le 4\arcmin$,
(c) $4\arcmin < R  \le 6\arcmin$, and (d) $9\arcmin <R \le 13\arcmin$.
The M105-SE field covers $2\arcmin \lesssim R \lesssim 6\arcmin$, while
the M105-W field covers $9\arcmin \lesssim R \lesssim 13\arcmin$.
We plotted a Hess diagram for the region with high number density in the CMDs to show better the details.
The dashed lines denote 50\% and 20\% completeness, and the dot-dashed line represents $M_{\rm bol}=-3.0$ mag.
%{\color{red}\bf
%Photometry of the resolved stars in the M105-W field %and NGC 3377 
%derived from the same $HST$ images was presented in \citet{har07b}. %, har07b}.
%However, our photometry derived from the images we produced goes about one magnitude deeper the red stars with $(V-I)=3.0$ than that in \citet{har07b} (see their Fig. 3), although both studies used the same raw $HST$ data. 
%The 50\% completeness magnitude for $(V-I)=3.0$ in our photometry (Figure 3(d)) is $I\approx 28.0$ mag, which is one magnitude fainter than that  in \citet{har07b}(their Fig.3), $I~27.0$ mag. 
%We infer that this difference is mainly due to the improvement of image combining in the drizzle package. }
For reference we overlayed theoretical RGB loci of the 12 Gyr isochrones for $[\alpha/{\rm Fe}] = 0.2$ and a range of metallicity ([Fe/H] = --2.4 to 0.2 with a step of 0.2 from left to right)
  in the Dartmouth models \citep{dot08}.
They were shifted according to the distance and foreground reddening of M105.
We adopted [M/H] = [Fe/H]+$[\alpha/{\rm Fe}]$, where $[\alpha/{\rm Fe}] = 0.2$.
%\citet{wei09} estimated from spectroscopic line indices that stellar populations at $2.6 - 3.5 R/R_{eff}$ are metal-poor and old (12 Gyr) {\color{red} {\bf (I think that the above sentence is not necessary.)}}. 

A few distinguishable features are noted in {\bf Figure \ref{fig_cmd}}.
First, most of the resolved stars in M105 are RGB stars. 
The RGB stars  show a large range of color, indicating that their metallicity ranges from very low ([Fe/H]$\approx -2.4$) to solar metallicity or higher. 
The sources brighter than the RGB stars appear to be a combination of old asymptotic giant branch (AGB) stars, foreground stars, and blended stars due to crowding, as discussed in the study of the M105-W field by \citet{har07b}.
%
%{\color{red}\bf
%Harris et al.(2007b) found about 40 stars brighter than the TRGB in the their CMD for the outer region. They estimated the expected number of foreground stars is about 20 and discovered 11 long period variable (PLV) candidates from the comparison of photometry for two epochs. Blending effect is considered to be negligible for these bright stars. They could not find any trace of younger AGB stars. They concluded that the presence of these bright stars can be explained as a combination of foreground stars and LPV stars in M105 and the fraction of 
%We also find about 45 stars with $24.5<I<25.6$ mag brighter than the TRGB  in our CMD for the outer region, which is consitent with the value given by Harris et al.(2007b). 
%However, we find a much larger number of these stars in the CMD of the inner region, N=311 at $3\arcmin<R<4\arcmin$ and N=219 at $4\arcmin<R<6\arcmin$. The number of the foreground stars for $24.5<I<25.6$ for each HST field is estimated to be 10, using TRILEGAL...
%AGB stars, blending  for the inner region....
%Results of the artificial star test: can tell the fraction of blending?????}
%
Second, 
the RGB  in the outer region ($9\arcmin<R \le 13\arcmin$) in {\bf Figure \ref{fig_cmd}(d)} 
shows clearly two distinguishable components: a narrow blue RGB and a wider red RGB, as shown also by \citet{har07b}.
The blue RGB is more clearly seen 
in this outer region than in the inner regions. 
%
%{%\color{red}\bf 
It is noted that our photometry of the outer region goes about one magnitude deeper for the red stars than that of the same region given by \citet{har07b}, although both studies used the same raw data. 
The 50\% completeness limit for $(V-I)=3.0$ is
$I \approx 28.0$ mag in our photometry, while it is $I\approx 26.8$ mag in
 \citet{har07b} (their {\bf Figure 3}). We infer that this difference is mainly due to the improvement of image combining in the drizzle package used in this study.
%}
%
Third, 
the positions of the isochrones for 12 Gyrs are consistent with the RGB in the CMDs, indicating that the RGB stars are mostly very old.

\begin{figure}
\centering
\includegraphics[scale=0.9]{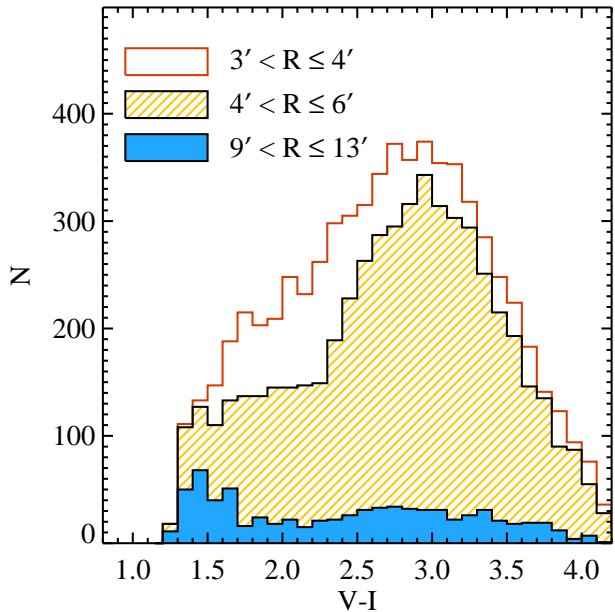} %{col_hist.eps} %{hist.eps} %{cmd_2.eps} %{testm105incmd1.eps} 
\caption{$(V-I)$ color distributions of the bright RGB stars 
($-3.7 < M_{\rm bol} \le -3.0$) at 
$3\arcmin< R  \le 4\arcmin$ (open histogram),
$4\arcmin < R \le 6\arcmin$ (hatched histogram), and 
$9\arcmin<R  \le 13\arcmin$ (filled histogram).
} 
\label{fig_cdf}
\end{figure}

In {\bf Figure \ref{fig_cdf}} we plotted the $(V-I)$ color distribution of the bright RGB stars ($-3.7 < M_{\rm bol} \le -3.0$) 
in the inner ($3\arcmin < R  \le 4\arcmin$ and $4\arcmin < R \le 6\arcmin$) and outer ($9\arcmin < R \le 13\arcmin$) regions.
We derived $M_{\rm bol}$ of the RGB stars as described in the next section.
The color distributions of the bright RGB stars in the outer region of  M105 show a clear bimodality:
a narrow blue component with a color range of $1.2 <(V-I)<1.7$ and  a peak at $(V-I) \approx 1.5$,
and a broad red component with a color range of   $1.8 <(V-I)<4.1$ and a peak at  $(V-I) \approx 2.8$.
The blue peak is higher than the red peak in this outer region.
The color distribution of the bright RGB stars in the inner region of  M105 shows also a bimodality, but in a weaker contrast than that of the outer region.

%%%%%%%%%%%%%%%%%%%%%%%%%%%%%%%%%%%%%%
%%%  Fig 4     
%%%%%%%%%%%%%%%%%%%%%%%%%%%%%%%%%%%%%%

\subsection{Distance Estimation}

%%%%%%%%%%%%%%%%%%%%%%%%%%%%%%%%%%%%%%
%%%  Fig 5    
%%%%%%%%%%%%%%%%%%%%%%%%%%%%%%%%%%%%%%
\begin{figure}
\centering
\includegraphics[scale=0.9]{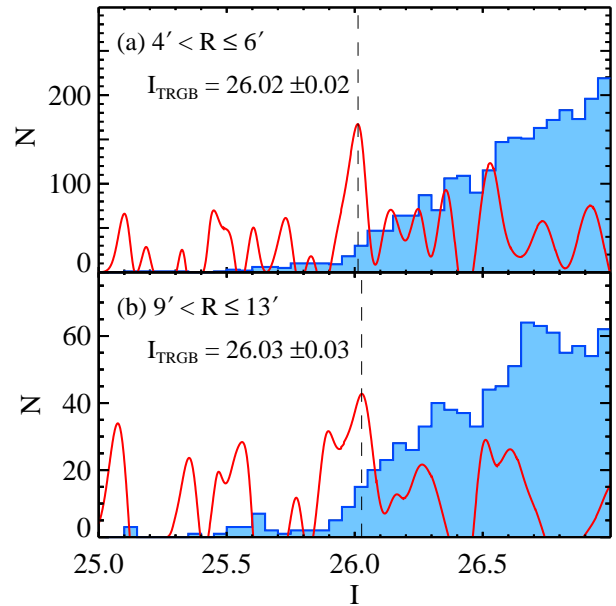} %{trgb.eps} 
\caption{$I$-band luminosity functions (histograms) and its corresponding edge detection response functions (red solid lines) for the red giants with $1.2<(V-I)\leq2.0$ 
at $4\arcmin<R <6\arcmin$  (a) and
 $9\arcmin<R <13\arcmin$ (b).
} 
\label{fig_ilf}
\end{figure}
 
We determined a distance to M105 using the TRGB method 
\citep{lee93,sak96,riz07} as follows. %,sal11}. %,con11}.
First, we derived the $I$-band luminosity functions of the red giants in the inner and outer regions ($4\arcmin<R \leq6\arcmin$ and  $9\arcmin<R \leq13\arcmin$) of M105 where crowding is not so severe.
The RGB stars show a broad range of color. 
We selected relatively blue stars with $1.2<(V-I)<2.0$ among the RGB stars for better measurement of the TRGB, and derived their $I$-band luminosity functions as shown in {\bf Figure \ref{fig_ilf}}.
%A jump at $I\approx 26.0$ mag in both regions  corresponds to the TRGB.

Using the edge-detecting algorithm \citep{men02, sak96} weighted according to the number of stars in each magnitude bin,
we determined the TRGB magnitude more quantitatively.
From the luminosity function $\Phi (m)$, we calculated an edge-detection response function 
$E(m)$ ($= \sqrt{\Phi}[ \log(\Phi (m + \sigma_m )) -   \log(\Phi (m - \sigma_m ))])  $,  
where %$\Phi (m)$ is the luminosity function of magnitude $m$ and 
$\sigma_m$ is the mean photometric error within a bin of $\pm0.05$ mag about magnitude $m$. 
%and we   weighted it  according to the Poisson noise of the luminosity function $E(m)\sqrt{\Phi (m) }$ \citep{men02}, 
They were plotted by the solid lines in {\bf Figure \ref{fig_ilf}}. %(b) and (d).
The errors for the TRGB magnitudes were 
obtained using the bootstrap resampling method. 
%We selected 1000 stars from the sample, and repeated it 1000 %times. As a results, we performed one million simulations. 
In each simulation we resampled randomly
the RGB sample with replacement to make a new sample of the same size. We performed one thousand simulations. We estimated the TRGB magnitude for each simulation using the same procedure, and derived the standard deviation of the estimated TRGB magnitudes.

Thus estimated TRGB magnitude for the inner region is %$I_{\rm TRGB} = 26.10\pm0.03$ and 
$I_{\rm TRGB} = 26.02\pm0.02$, which is almost the same as that for the outer region,  $I_{\rm TRGB} = 26.03\pm0.03$.
The median color of the TRGB is derived from the colors of the selected bright red giants close to the TRGB:
 $(V-I)_{\rm TRGB} =  1.61\pm0.02$ for the inner region, and
 $1.54\pm0.03$ for the outer region.
We used the TRGB calibration given in \citet{riz07}: 
$M_{{\rm I,TRGB}} = -4.05(\pm0.02) + 0.217(\pm0.01)( (V-I)_0 -1.6)$ (where $(V-I)_0$ is a reddening corrected color of the TRGB) with a systematic error of 0.12.
%,  which  is very similar to that given in \citet{tam08}. %, $M_{{\rm I,TRGB}} = -4.05\pm+ 0.02$ for $(V-I)_0 =1.6$
%Then we calculate the distance modulus using $(m-M)_0 = I_{\rm 0, TRGB} - M_{\rm I, TRGB}$.
After correction for foreground reddening, we derived a distance modulus 
$(m-M)_0=30.04\pm0.02$ for the inner region, and
 $30.07\pm0.03$ for the outer region.
 We adopted a mean of these two estimates,
$(m-M)_0=30.05\pm0.02$ with a systematic error of 0.12
(corresponding to a linear distance of $10.23\pm0.09$ Mpc).
At this distance one arcsec corresponds to 49.6 pc.
The results of distance estimation for M105 
are summarized in {\bf Table \ref{tab_distance}}. 
This shows that M105 is located at the distance similar to those of M66, M96, and NGC 3377 in the same Leo I Group \citep{lee13}.
%This value for M105 is in excellent agreement  with the result derived from the M105-W Field data
%by \citet{har07b}, $(m-M)_0=30.06\pm0.10$.

%Its systematic error is estimated to be 0.12, considering (a) the TRGB calibration error of 0.12 \citep{bel04, mag08},
%(b) the aperture correction error of 0.02, (c) and the standard calibration error of 0.02 for ACS \citep{sir05}.

%\subsection{Comparison of Distance Estimates for M105}

%%TRGB ===========
%This study:                 30.10 0.16 10.5 Mpc
%\citet{har07b} TRGB 30.10 0.16 10.5 Mpc
%\citet{gre04} TRGB(K) 30.14 0.14 10.7 Mpc
%\citet{sak97} TRGB    30.30 0.14  11.5 Mpc
%%SBF ===========
%\citet{jen03} SBF 29.96  0.11 9.82 Mpc
%\citet{sil01} SBF 30.12 0.11 10.6 Mpc
%%PNLF ===========
%\citet{cia02} PNLF 29.90  0.11 9.55 Mpc
%%\citet{93} PNLF 30.01 0.09 10.0 Mpc

%%%%%%%%%%%%%%%%%%%%%%%%%%%%%%%%
% Table 3
%%%%%%%%%%%%%%%%%%%%%%%%%%%%%%%%

\begin{deluxetable*}{lcc}
%\tabletypesize{\footnotesize} % \footnotesize \scriptsize
%\tabletypesize{\tiny}
\setlength{\tabcolsep}{0.05in}
%\rotate
\tablecaption{A Summary of TRGB Distance Measurements for M105}
\tablewidth{0pt}

\tablehead{ \colhead{Parameter}  &  \colhead{M105-SE} & \colhead{M105-W} 
} 
\startdata
TRGB magnitude, $I_{TRGB}$					&  $26.02\pm0.02$ & $26.03\pm0.03$ \\
TRGB color, $(V-I)_{TRGB}$					&  $1.61\pm0.02$ & $1.54\pm0.03$ \\
Foreground extinction at $V$, $A_{V}$	& 0.071 & 0.062\\
Foreground extinction at $I$, $A_{I}$	&  0.039 & 0.034\\
Foreground reddening, $E(V-I)$		& 0.032 & 0.028 \\
Intrinsic TRGB magnitude $I_{0,TRGB}$		& $25.98\pm0.03$ & $26.00\pm0.03$\\
Intrinsic TRGB color, $(V-I)_{0,TRGB}$		&  $1.58\pm0.03$ & $1.51\pm0.03$\\
Absolute TRGB magnitude, $M_{I,TRGB}$		&$-4.06$ & $-4.07$\\
Distance modulus, $(m-M)_0$					& $30.04\pm0.02$ & $30.07\pm0.03$\\
                          & \multicolumn{2}{c}{mean = $30.05\pm0.02\pm0.12$ }  \\
Distance					&  $10.17\pm0.09$ Mpc & $10.30\pm0.14$ Mpc\\
					& \multicolumn{2}{c}{mean = $10.23\pm0.08$ Mpc}  \\
\hline
%\tablenotetext{a}{$(m-M)_{0,M66}=30.12\pm0.03$ (Lee \& Jang 2013)}
%\tablenotetext{b}{$(m-M)_{0,M96}=30.15\pm0.03$ (Lee \& Jang 2013)}
%\tablenotetext{c}{Angular separation between NGC 3384 and M105 : 0.1207 deg}
%\tablenotetext{d}{Three dimensional distance between NGC 3384 and M105 : $\sqrt{ 163?? ^2 + 380^2 }$ = 400?? kpc }
%\tablenotetext{e}{Angular separation between NGC 3384 and M96 : 0.89 deg}
%\tablenotetext{f}{Three dimensional distance between NGC 3384 and M96 : $\sqrt{ (163 kpc)^2 + (250 kpc)^2 }$ = 298 kpc }
\enddata
\label{tab_distance}
\end{deluxetable*}

The distance to M105 was estimated using various methods in the past.
Our estimate based on the TRGB method ( $(m-M)_0 = 30.05\pm 0.02({\rm ran})\pm0.12 ({\rm sys})$)  is consistent with those derived using the planetary nebula luminosity function 
($30.09\pm 0.09$ \citep{cia93}, $29.98\pm 0.11$ \citep{cia02}), and
the surface brightness fluctuation method 
( $30.32\pm0.14$  \citep{ton01}, and   $30.16\pm0.14$ \citep{jen03}). 
 However, our value is much smaller than those based on the globular cluster luminosity function. \citet{lar01} derived $30.93\pm 0.12$ from the  total globular cluster sample, $31.92\pm 0.30$ from the red globular cluster sample, and $30.15\pm 0.12$ from the blue globular cluster sample. The first two values are much larger than ours, while the last is consistent with ours. However, these estimates were based on a small number of globular clusters (30 blue globular clusters and 24 red globular clusters with $20<V<24$ mag) so that their estimates must suffer from a significant uncertainty.
 
There are a few studies of the TRGB distance measurements for M105 in the literature. 
\citet{sak97} presented TRGB measurements,
%$I_{TRGB}=26.42\pm0.09$ and $(m-M)_0=30.42$ 
$I_{TRGB}=26.32\pm0.05$ and $(m-M)_0=30.30\pm0.14$ 
(adopting $A_I = 0.05$), from the $HST$/WFPC2 images.
Their TRGB magnitude is 0.3 mag fainter than our measurement, $I_{\rm TRGB} = 26.03\pm0.03$. It is noted that \citet{sak97} used only F814W images
for their analysis, while we used much deeper F606W and F814W images.
\citet{gre04} also derived a distance estimate from the F160W band TRGB magnitude based on the HST/NICMOS images,
$(m-M)_0=30.14\pm0.14$. 
Later \citet{har07b} presented TRGB measurements based on the same HST/ACS images of the M105-W field as used in this study,
$I_{TRGB}=26.10\pm0.10$ and $(m-M)_0=30.06\pm0.10$.
Thus the values for the TRGB magnitude and distance in this study, \citet{har07b}, and \citet{gre04} are in excellent agreement, but these TRGB values are 0.2 to 0.3 mag brighter than the value given by \citet{sak97}. 

%The reason for this difference is not clear. However, it it noted that our value and the \citet{har07b} value agree within 0.1 mag, and this indicates our estimate is more reliable than that given by \citet{mou09}.\\
%{\color{red} \bf $<$- There is no TRGB distance estimate of NGC 3379 in \citet{mou09}! TRGB distance estimates in previous studies are summarized in Table \ref{tab_trgb}}
%that our measurement more reliable than \citet{mou09}'s. 
%\clearpage

%%%%%%%%%%%%%%%%%%%%%%%%%%%%%%%%%%%%%%
%%%  Fig 6 New 2015 Dec
%%%%%%%%%%%%%%%%%%%%%%%%%%%%%%%%%%%%%%
\begin{figure}
\centering
\includegraphics[scale=0.9]{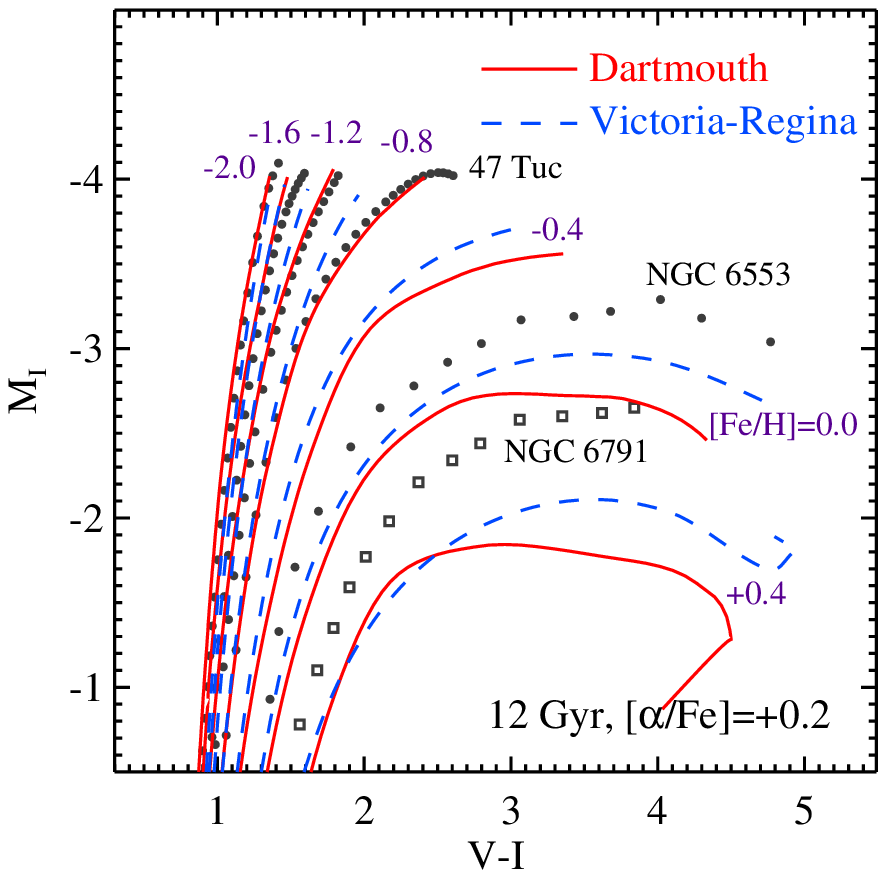} %isochrone.eps} %{mdf0.eps} %{mdf1.eps} %{mdf_n3384.eps} 
\caption{RGBs of 12 Gyr isochrones for [$\alpha$/Fe]=+0.2 and [Fe/H]=--2.0, --1.6, --1.2, --0.8, --0.4, 0.0, and +0.4 in the Dartmouth models \citep{dot08} (solid lines) and in the Victoria-Regina models \citep{van14} (dashed lines).
Dotted lines represent the RGB sequences of five galactic globular clusters \citep{dac90,har96}:
M15 ([Fe/H]$=-2.37$),
M2 ([Fe/H]$=-1.65$),
NGC 1851 ([Fe/H]$=-1.18$),
47 Tuc (NGC 104) (Fe/H]$=-0.72$), and
NGC 6553 ([Fe/H]$=-0.18$),
Open squares denote the RGB sequence of NGC 6791, an  old open cluster with [Fe/H]$=+0.3$ and age of 8.5 Gyr \citep{gar94,cha99,van14}. 
}
\label{fig_iso}
\end{figure}

\subsection{MDFs  of the RGB Stars}
%Metallicity Distribution Functions

We estimate the metallicity ([M/H]) of the bright RGB stars in M105, comparing their $(V-I)$ colors with the theoretical isochrones, following the method used in \citet{har07a,har07b}. 
%We used the 12 Gyr isochrones for [$\alpha$/H]$=0.2$ and a range of metallicity %([Fe/H] = --2.4 to +0.2) 
%([M/H] = --2.2 to +0.4)   in the Dartmouth models \citep{dot08}.
%, as plotted in Figure \ref{fig_cmd}.
We assumed that the mean age of the stars in this galaxy is 12 Gyr, as found from 
spectroscopic measurement of mean ages of the stellar populations in  the regions at 2.6--3.5 $R/R_{\rm eff}$ of M105 by \citet{wei09}.
%that of typical globular clusters, 12 Gyrs, as in \citet{har07a,har07b}. 

%{%\color{red}
%\bf 
In {\bf Figure \ref{fig_iso}} we plotted the CMDs for the RGBs of 12 Gyr isochrones 
%with [$\alpha$/Fe] = +0.2 and [Fe/H] = --2.0, --1.6, --1.2, --0.8, --0.4, 0.0, and +0.4
 in the Dartmouth models \citep{dot08} in comparison with one of the recent models for old stellar populations, the Victoria-Regina models \citep{van12, van14}.
We also overlayed the RGB sequences of five Milky Way globular clusters \citep{dac90}: 
M15 ([Fe/H] $=-2.37$),
M2 ([Fe/H] $=-1.65$),
NGC 1851 ([Fe/H] $=-1.18$),
47 Tuc (NGC 104, [Fe/H] $=-0.72$),
NGC 6553 ([Fe/H] $=-0.18$),
and one old open cluster NGC 6791 ([Fe/H] $=+0.2$ to +0.35) \citep{gar94,cha99}. %, as shown also in \citet{har00}.
The [Fe/H] values of the globular clusters are from the 2010 edition of  the catalog of the Milky Way globular clusters \citep{har96}.  
NGC 6791 is known to be 8.5 Gyr old, a few Gyrs younger than the globular clusters, and its metallicity is estimated to be [Fe/H]$=+0.20$ to +0.35 \citep{har00,van14}.

A comparison of the Dartmouth isochrones with the RGBs of the globular clusters shows that the isochrones for [Fe/H] = --2.0, --1.6, --1.2, and --0.8 are roughly  in agreement with the RGBs of M15, M2, NGC 1851, and 47 Tuc, respectively.
The RGB of the metal-rich bulge globular cluster NGC 6553 with [Fe/H] $=-0.18$ is seen between the Dartmouth isochrones for [Fe/H] = --0.4 and [Fe/H] = $0.0$, and that of the old open cluster NGC 6791  is located between the isochrones for [Fe/H] = 0.0 and [Fe/H] = $+0.4$.
On the other hand, 
%a comparison of the two models finds that 
the Victoria-Regina isochrones are slightly redder for [Fe/H] = --2.0, similar for [Fe/H] = --1.6, and bluer or brighter than for [Fe/H] $>-1.6$, compared with the Dartmouth isochrones.

The RGBs of the isochrones become redder (cooler) as the value of [$\alpha$/Fe] increases. 
%Mean  [$\alpha$/Fe] values are +0.3 to +0.4 for metal-poor globular clusters, and they decrease as the metallicity increases.
We adopted, as a reference value, [$\alpha$/Fe] = +0.2, for the study of the resolved stars with a large range of metallicity in M105.
In this study we used the Dartmouth isochrones as provided.
It is noted that recognizing the difference between the old versions of Victoria-Regina models and the RGB sequences of clusters, 
\citet{har00,har02}  adjusted slightly the colors of the isochrones using the cluster sequence for their study of the MDFs. 

%Procedures for deriving MDFs
We derived a metallicity of each star in M105 from the comparison of the CMDs with the Dartmouth isochrones.
The problem due to the curvature of the RGBs for high metallicity is more serious in the ($M_I$, $(V-I)_0$) plane compared with in the ($M_{\rm bol}$, $(V-I)_0$) plane so some of the previous studies adopted the latter plane for interpolation \citep{har00,har02}. % latter plane.
However, it is not easy to convert an observational ($M_I$, $(V-I)_0$)  plane of stars to an ($M_{\rm bol}$, $(V-I)_0$) plane for the metal-rich stars because of the significantly curved RGB sequences, while 
it is straightforward to convert a theoretical ($M_{\rm bol}$, $(V-I)_0$) plane of models to an  ($M_I$, $(V-I)_0$) plane. 
Therefore we used ($M_I$, $(V-I)_0$) planes rather than ($M_{\rm bol}$, $(V-I)_0$) planes for interpolation to derive a metallicity value of each star.
To reduce the uncertainties due to the curved tracks and non-linearity in the [M/H]-color relations as much as possible, we generated
fine grids with steps d[M/H] = 0.01 (for [M/H]$>-0.4$), 0.02  (for $-0.6<$[M/H]$\leq -0.4$), and 0.1 (for $-2.4<$[M/H]$\leq -0.6$) using the Dartmouth models.  
For this we used non-linear interpolation between the tracks.
Thus the errors in interpolations are supposed to be negligible. 
%}

The color range for the same metallicity range becomes wider, as the RGB stars become brighter, as seen in {\bf Figure \ref{fig_cmd}}. 
In addition,
the photometric errors become smaller and the photometric completeness gets higher, as the stars become brighter. 
Therefore the derived MDFs are considered to be more reliable for brighter RGB stars. 
The $I$-band magnitudes of the reddest (the highest metallicity) stars become fainter as the color increases so that some of them may be missed, being affected by the incompleteness of our photometry. Therefore 
we derived the metallicity of the stars in the %$M_{\rm bol} - (V-I)_0$ 
$M_I - (V-I)_0$ diagram where $(V-I)_0$ represents colors corrected for foreground reddening.

\begin{figure}
\centering
\includegraphics[scale=1.0]{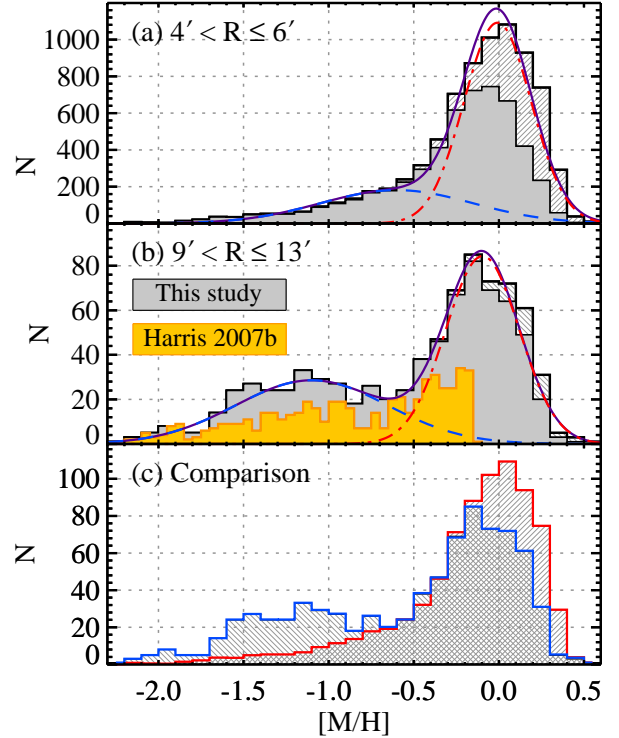} %{mdf3.eps} %{fig6.eps} %{mdf0.eps} %{mdf1.eps} %{mdf_n3384.eps} 
\caption{(a,b) MDFs for the bright RGB stars ($-3.7 < M_{\rm bol}\le -3.0$) in the inner region ($4\arcmin<R \le 6\arcmin$) and the outer region ($9\arcmin<R \le 13\arcmin$) of M105. 
%Upper and lower panels represent the $M_{\rm bol}$ range of  $-3.7 < M_{\rm bol}\le -3.0$, and 
%(a) $-3.0 < M_{\rm bol}\le -2.5$, respectively.
Filled histograms and hatched histograms show 
the MDFs before and after completeness correction, respectively.
%the observed MDFs, while hatched histograms show the MDFs corrected for completeness of photometry.
Blue dashed lines, red dot-dashed lines, and purple solid lines represent the double Gaussian fits for completeness-corrected MDFs: metal-poor component, metal-rich component, and their sum.
The yellow histogram represents the completeness-corrected MDF of the bright RGB stars ($-3.5<M_{\rm bol} <-3.0$) in the outer region given by \citet{har07b}(their {\bf Figure 6}). Note that their bin size is half of ours so that their numbers in each bin are about half of ours.
(c) A comparison of the MDFs of the inner region (red  histogram) and the outer region (blue histogram). The MDF of the inner region is scaled to match that of the outer region at [M/H] $=-0.55$. Note the excess of low metallicity stars in the outer region above the MDF of the inner region.} 
\label{fig_mdf}
\end{figure}

In {\bf Figure \ref{fig_mdf}} we display the MDFs for the bright RGB stars (with the bolometric magnitude $M_{\rm bol} \leq -3.0$ mag) in the inner  region ($4\arcmin<R \leq 6\arcmin$) and the outer region ($9\arcmin<R  \leq 13\arcmin$) of M105 where the completeness of our photometry is the highest.
For deriving the MDFs we used the regions where photometric completeness is higher than 50\% for $M_{\rm bol} = - 3.0$ and [M/H]$<0.2$, selecting a less crowded region at $R>4\arcmin$ in the case of the inner field. %We derived the MDFs down to $M_{\rm bol} = - 3.0$. %, and corrected them for completeness.
We corrected the incompleteness of the MDFs using the completeness test results derived in this study. Completeness correction is negligible in the outer region, while it is significant in the high end of the MDFs in the inner region. However, it is noted that the peak position of the metal-rich component increased only by about 0.1 dex after completeness correction even in the inner region.
%two bins of  bolometric magnitudes, $-3.7<M_{\rm bol} \leq -3.0$  and $-3.0 < M_{\rm bol} \leq -2.5$.

%%%%%%%%%%%%%%%%%%%%%%%%%%%%%%%%%%%%%%
%%%  Fig 7
%%%%%%%%%%%%%%%%%%%%%%%%%%%%%%%%%%%%%%

%%%%%%%%%%%%%%%%%%%%%%%%%%%%%%%%
% Table 4
%%%%%%%%%%%%%%%%%%%%%%%%%%%%%%%%

\begin{deluxetable*}{lcccccc}
\tabletypesize{\footnotesize} % \footnotesize \scriptsize
%\tabletypesize{\tiny}
\setlength{\tabcolsep}{0.05in}
%\rotate
\tablecaption{A Summary of Gaussian Fits for MDFs of Bright RGB stars in M105}
\tablewidth{0pt}

\tablehead{ \colhead{Region} %& Magnitude 
&  \multicolumn{3}{c}{Metal-poor component} & \multicolumn{3}{c}{Metal-rich component} \\
& Center & Width & $N_{\rm total}$ & Center & Width & $N_{\rm total}$ 
} 
\startdata
%\multirow{2}{*}{$4' < R_{GC} \leq 6'$} 	%&  $-3.7 < M_{bol} \leq-3.0$ 
$4' < R \leq 6'$ & $-0.60\pm0.04$ & $0.45\pm0.03$ & $2040\pm218$ &  $-0.01\pm0.01$ & $0.20\pm0.01$ & $5405\pm220$ \\
%										&  $-3.0 < M_{bol} \leq-2.5$ 
%& $-0.58\pm0.05$ & $0.57\pm0.03$ & $3975\pm299$ &  $0.12\pm0.01$ & $0.23\pm0.01$ & $4673\pm274$ \\
%\multirow{2}{*}{$9' < R_{GC} \leq 13'$}	 %&  $-3.7 < M_{bol} \leq-3.0$ 
$9' < R \leq 13'$ & $-1.10\pm0.05$ & $0.45\pm0.05$ & $320\pm29$ &  $-0.10\pm0.02$ & $0.21\pm0.01$ & $457\pm31$\\
%										&  $-3.0 < M_{bol} \leq-2.5$ 
%& $-1.09\pm0.05$ & $0.52\pm0.04$ & $581\pm42$ &  $-0.03\pm0.03$ & $0.23\pm0.02$ & $324\pm37$\\
\hline
\enddata
\label{tab_fits}
\end{deluxetable*}

{\bf Figure \ref{fig_mdf}} shows two important features as follows.
%{%\color{red}
%\bf 
First, the MDF of the outer region is clearly bimodal, showing two distinguishable  components, while the visibility of the metal-poor component is much weaker in the inner region.
To show this better, we scaled the MDF of the inner region to match that of the outer region at
[M/H] = --0.55, and overlayed the scaled result in {\bf Figure \ref{fig_mdf}(c)}. It is seen clearly that there is a significant excess above the MDF of the inner region in the low metallicity range at [M/H] $<-0.7$. 
This is distinct from typical forms of the MDFs of stars in massive galaxies that show a single major metal-rich component with a long weak tail in the low metallicity \citep{har02}. 
This is also in strong contrast against the MDF of the same region given by \citet{har07b} (their {\bf Figure 6}), which is overlayed by the yellow histogram in {\bf Figure \ref{fig_mdf}(b)}. \citet{har07b} pointed out that the MDF of the outer region in M105 is the broadest and flattest among the known MDFs of galaxies, and that they could not find any dominant subpopulation in their MDF. 

To check and quantify the bimodality, we fit the data with two Gaussian functions using mpfitexpr task in IDL, as summarized in {\bf Table \ref{tab_fits}}. % and \ref{tab_fits2}.
%(TO Add GMM test results?????!!!!!!!
We tested also the bimodality of the MDFs of these stars using the Gaussian Mixture Modelling (GMM) package program provided by \citet{mur10}.
%, summarizing the results in Table \ref{tab_cdfgmm}.
%The probability for the unimodal distribution ($P$) and the difference of two peaks ($D$) are two useful indicators  when testing the bimodality of the data. The condition, $P<0.1\%$, is used for non-unimodal distributions, and another condition, $D>2$, is for clear bimodality. We adopted a homoscedastic case (same variance) for bimodality.
These tests show that a unimodal distribution is rejected for both the inner and outer regions (the probability for the unimodal distribution $p$ is much smaller than 0.1\%). The MDF for the outer region is clearly bimodal with peaks at [M/H] $=-0.09$ and --1.12 (the value of the bimodality parameter, $D=2.93\pm0.24$,  is larger than the bimodality criterion 2), while that for the inner region is marginally bimodal with peaks  at [M/H] $=-0.01$ and --0.68 
(the value of $D$, $1.78\pm0.03$, slightly smaller but close to 2).
The MDFs are reasonably well fit by the double Gaussian functions not only for the outer region but also for the inner region. 
However, this Gaussian function fitting is only for showing the presence of bimodality in the MDFs, without any physical justification.
We will discuss these MDFs further in terms of chemical evolution models in Section 4.1.
%}

%%%%%%%%%%%%%%%%%%%%%%%%%%%%%%%%%%%%%%
%%%  Fig 8
%%%%%%%%%%%%%%%%%%%%%%%%%%%%%%%%%%%%%%

%%%%%%%%%%%%%%%%%%%%%%%%%%%%%%%%%%%%%%
%%%  Fig 8
%%%%%%%%%%%%%%%%%%%%%%%%%%%%%%%%%%%%%%
\begin{figure*}
\centering
\includegraphics[scale=0.85]{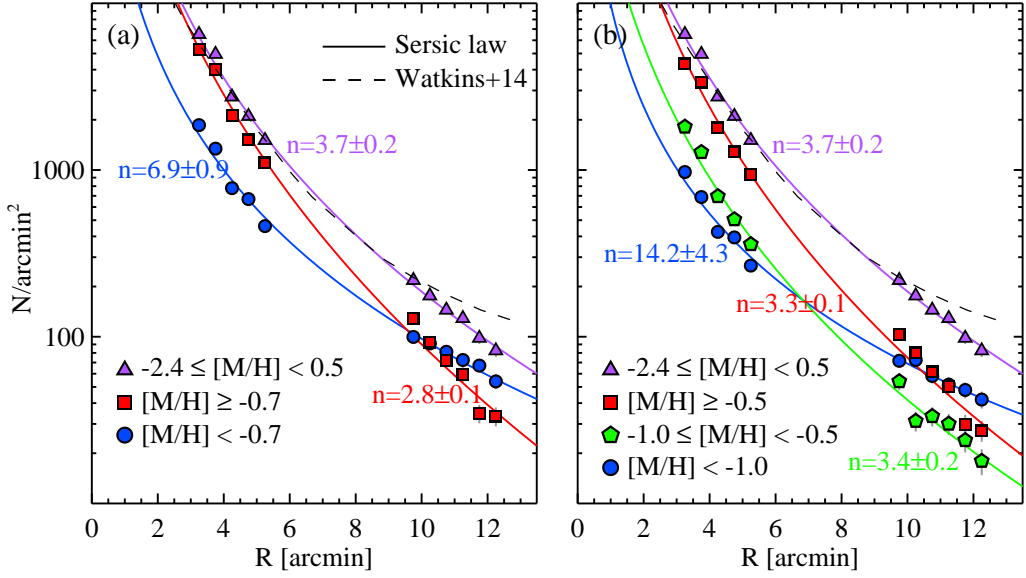} %{mdf3.eps} %{fig6.eps} %{mdf0.eps} %{mdf1.eps} %{mdf_n3384.eps} 
\caption{ Radial number density profiles of the bright RGB stars ($I \leq 27.0$) in M105 corrected for completeness.
(Left panel) Circles, squares, and triangles represent the metal-poor stars (${\rm [M/H]} \leq -0.7$),  metal-rich stars (${\rm[M/H]} > -0.7$), and all stars,  respectively.
(Right panel) Circles, pentagons, squares, and triangles represent the metal-poor stars (${\rm [M/H]} \leq -1.0$),  intermediate metallicity stars ($-1.0<{\rm[M/H]} \leq -0.5$), metal-rich stars (${\rm[M/H]} > -0.5$), and all stars,  respectively.
The dashed line represents the $V$ band surface brightness profile along the major axis in
%\citet{dev79, cap90}. 
\citet{wat14}.
Solid lines represent Sersic profile fits of each stellar number density profile.
% : $n=2.8\pm0.8$ for metal poor stars, $n=1.9\pm0.3$ for metal rich stars, and $n=2.2\pm0.4$ for all stars.
%{Choose left, right, or both???????????}} 
}
\label{fig_radden}
\end{figure*}

%%%%%%%%%%%%%%%%%%%%%%%%%%%%%%%%%%%%%%
%%%  Fig 9
%%%%%%%%%%%%%%%%%%%%%%%%%%%%%%%%%%%%%%
\begin{figure*}
\centering
\includegraphics[scale=0.85]{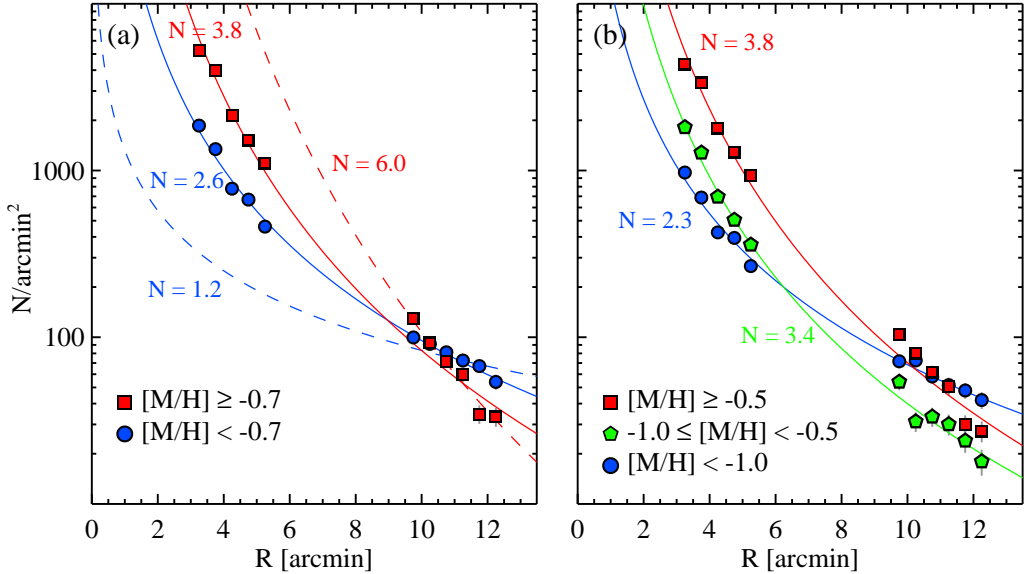} %{mdf3.eps} %{fig6.eps} %{mdf0.eps} %{mdf1.eps} %{mdf_n3384.eps} 
\caption{ Same as {\bf Figure \ref{fig_radden}}, but for the power law fits. 
%Solid lines represent the power law fits with an index $N=2.58\pm0.03$ for metal-poor stars and an index $N=3.83\pm0.03$ for metal rich stars.
Dashed lines represent the power law fit results that \citet{har07b} derived for the outer region ($9\arcmin<R<13\arcmin$) only:
$N=1.2\pm0.7$ for the metal-poor stars and $N=6.0\pm0.6$ for the metal-rich stars. Note that they do not match the data for the inner region.
}
\label{fig_radden2}
\end{figure*}

Second, the peak metallicity of the metal-rich component in the inner region is [M/H] $= -0.01\pm0.01$, %0.05$, very
which is slightly higher than %similar to 
that for the outer region, [M/H] $= -0.10\pm0.02$. %05$.
In contrast, the peak metallicity of the metal-poor component in the inner region is [M/H] $= -0.60\pm0.04$, somewhat higher than that for the outer region, [M/H] $= -1.10\pm0.05$.
%{%\color{red} \bf 
However, typical MDFs of stars in the inner regions of massive galaxies are not Gaussian, but follow a 
dominant quasi-gaussian form with a long tail toward the low metallicity end. Therefore the value for the genuine peak of the metal-poor component in the inner region must be much lower than [M/H] $=-0.6$, which will be shown in Section 4.1.
%}

\subsection{Radial Distributions  of the RGB Stars}

Since the MDFs of the RGB stars are bimodal, we divided the sample of bright RGB stars %($I \leq 27$ mag) 
into two groups according to their metallicity: %with $[{\rm M/H}]=-0.7$ :
 a metal-poor group ($[{\rm M/H}]\leq -0.7$),  and a metal-rich group ($[{\rm M/H}]>-0.7$).
%{\color{red}\bf TO REMOVE or KEEP? into three groups: 
% a metal-poor group ($[{\rm M/H}]\leq -1.0$), an intermediate metallicity group ($-1.0<[{\rm M/H}]<-0.5$), and a metal-rich group ($-0.5<[{\rm M/H}]$).}, 
We selected a cut with $[{\rm M/H}]=-0.7$, for which the MDF shows the minimum value between the two peaks in M105.
Then we derived radial number density profiles of the bright stars ($I\leq 27.0$ mag) in these groups, 
and plotted them in {\bf Figure \ref{fig_radden}(a)}.
We corrected these profiles using the completeness data we derived in Section 2.2. 
We also plotted the $V$-band surface brightness profile (solid lines) of M105 given by \citet{wat14}  (in their {\bf Figure 6}) for comparison.
This surface brightness profile extends out to the semi-major axis radius, SMA$\approx 850\arcsec$, covering a wider area than the old data given in \citet{cap90}.
We fit the radial number density profiles with
a Sersic law \citep{ser63} and a power law, adopting no weighting.

Several distinguishable features are noted  in this figure as follows.
First, the radial number density profile of the metal-rich stars is described by a Sersic law with $n=2.75\pm0.10$  %$n=1.5\pm0.2$ 
for the entire range of the galactocentric radius ($3\arcmin < R < 13\arcmin$), 
while that of the metal-poor stars  is fit well by
a Sersic law with $n=6.89\pm0.94$. %$n=2.0\pm0.5$. %, similar to the de Vaucouleur law.
The radial number density profile of the total sample is fit well with a Sersic index $n=3.71\pm0.21$, close to the value of the de Vaucouleurs law. %$n=1.8\pm0.2$.
%{\color{red}\bf ($1.8\pm0.1$???? in the new fitting)}.
% with $n=4$ that fits well the surface brightness profile.
Second, the slopes of the radial number density profiles of both groups and the surface brightness profile are similar in the inner region.
However, the slope of the radial number density profiles of the total sample is much steeper than that of the surface brightness profile in the outer region where the surface brightness is too faint to be measured reliably \citep{wat14}.
Third,
the radial number density profile of the metal-poor stars is much flatter than that of the metal-rich stars in the outer region.
%{\color{red} \bf (Not n=4)}. %, with a power law index of $??\pm ??$.
Thus the radial number density profile of the metal-poor stars is more extended than that of the metal-rich stars in the outer region.
Fourth, the mean number density of the metal-rich stars is much larger (several times) than that of the metal-poor  stars in the inner region. The number density of the metal-poor  stars become similar to that of the metal-rich stars in the outer region at $R\approx 10\arcmin$, and exceed the latter beyond  $R\approx 10\arcmin$.
%{\color{red}
Fifth, 
%we estimate the effective radii for both groups, obtaining $R_{\rm eff} = 1\farcm03\pm0\farcm03$ for the metal-poor group, and $R_{\rm eff} = 0\farcm38 \pm0\farcm03$ for the metal-rich group. 
%Thus the effective radius of the metal-poor group is much larger than that of the metal-rich group.
The effective radius for the total sample is derived to be $R_{\rm eff} = 0\farcm92 \pm 0\farcm10$, which is consistent with the effective radius of the integrated stellar light of the galaxy, $58\farcs7$. 
%Thus the effective radius of the metal-poor group is much smaller than that of the metal-rich group. 
%What does it mean????
%It indicates that the origin of the metal-poor stars is two:
%one in the inner region from the in situ formation, and the other in the outer region from the accretion of satellite dwarf galaxies.....
%}

\citet{har07b} found that the radial profile of the metal-rich 
([M/H]$>-0.7$) stars in the M105-W Field is fit well with a power law, 
$\sigma \sim R^{-6.0\pm0.6}$, much steeper  than that of the metal-poor  ([M/H]$<-0.7$) stars, $\sigma \sim R^{-1.2\pm0.7}$.
We plotted these power laws by the dashed lines in {\bf Figure \ref{fig_radden2}(a)}.
%\\ {\bf To modify}\\
The radial density profiles of the outer region derived in this study are consistent  with these two power laws.
However, those of the inner region show large deviations from these two power laws: the radial profile of the metal-rich stars is flatter 
than, and that of the metal-poor stars is steeper than the values extrapolated from the \citet{har07a}'s fits. 
The radial profiles of the metal-rich stars and metal-poor stars in the entire range are fit well by power laws with $\sigma \sim R^{-3.83\pm0.03}$ and $\sigma \sim R^{-2.58\pm0.03}$, respectively.

%{%\color{red}
%\bf
After subtracting the metal-rich component derived from the accreting gas model of chemical evolution in Section 4.1, the residual MDFs of the metal-poor components in the inner region and the outer region are found to be similar, as shown in {\bf Figure 10(c)}.  A cut with [M/H]$<-0.7$ covers 93\% of the metal-poor component in the inner region, and a similar fraction, 95\%, in the outer region. %On the other hand, 
Thus the contribution of the low metallicity tail of the metal-rich component to this range is  6.8\% in the inner region and 8.6\% in the outer region. 
Thus there may be some selection bias in the derived radial profiles with a given [M/H] cut value. 
To check this, we divided the [M/H] range into three groups [M/H] $<-1$, (--1 to --0.5), and $>-0.5$, and derived the radial profiles for each, as shown in the right panels of {\bf Figures 8} and {\bf 9}. 
In this case, the range [M/H] $<-1.0$ covers 55\% of the metal-poor component in the inner region, and 76\% in the outer region. The contribution of the low metallicity tail of the metal-rich component to this range is 3.0\% in the inner region and 3.5\% in the outer region.
It is expected that the radial profiles of the metal-poor component for [M/H] $=-1.0$ cut will be flatter than that for [M/H] $=-0.7$ cut.
This is seen in {\bf Figures 8(b)} and {\bf 9(b)}: the power index of the metal-poor component for [M/H]  $=-0.7$ cut is $N=2.58\pm0.03$, which is slightly larger than the value for [M/H] $=-1.0$ cut, $N=2.28\pm0.02$. %?\pm??$. 
On the other hand, the value of the power law index for the metal-rich component does not change depending on the [M/H] cut value, with $N=3.83\pm0.03$ for [M/H] $\geq -0.7$ and $N=3.82\pm0.02$ for [M/H] $\geq -0.5$. Thus there is some selection bias in the radial profiles due to the [M/H] cut value, but it is not significant.

\section{Discussion}

%{%\color{red} 
%\bf

\subsection{Analytic Chemical Evolution Models for the MDFs}

%%%%%%%%%%%%%%%%%%%%%%%%%%%%%%%%
% Table 5
%%%%%%%%%%%%%%%%%%%%%%%%%%%%%%%%
\begin{deluxetable*}{lcccccc}
\tabletypesize{\footnotesize} % \footnotesize \scriptsize
%\tabletypesize{\tiny}
\setlength{\tabcolsep}{0.05in}
%\rotate
\tablecaption{A Summary of Chemical Evolution Model Fits}
\tablewidth{0pt}

\tablehead{ 
\colhead{Region} & \colhead{Component} &  \colhead{Pristine} & \multicolumn{2}{c}{Pre-enriched} &  \multicolumn{2}{c}{Accreting gas} \\
\colhead{} & \colhead{} &  \colhead{closed-box} & \multicolumn{2}{c}{closed-box} &  \multicolumn{2}{c}{} \\
	& & $p(Z_\odot)$ & $p(Z_\odot)$ & [M/H]$_0$ & $p(Z_\odot)$ & $M$  
} 
\startdata
Inner region & Single & $0.84\pm0.01$ & $0.81\pm0.01$ & $-2.18\pm0.03$ & $1.14\pm0.02$ & $2.93\pm0.18$ \\
Outer region & Single & $0.62\pm0.03$ & $0.62\pm0.03$ & $\leq-4.0$ & $0.79\pm0.02$ & $1.36\pm0.51$ \\
\hline
               & Double &  &  &  &  &  \\
Inner region& Metal-rich &  &  &  & $1.14\pm0.02$ & $2.93\pm0.18$ \\
			& Metal-poor & $0.11\pm0.01$ & $0.10\pm0.01$ & $-2.42\pm0.03$ & $0.10\pm0.02$ & $1.35\pm0.63$ \\
Outer region& Metal-rich & &  &  & $0.90\pm0.04$ & $2.84\pm0.51$ \\
			& Metal-poor & $0.06\pm0.01$ & $0.05\pm0.01$ & $-2.29\pm0.01$ & $0.06\pm0.01$ & $1.83\pm0.56$ \\

\enddata
%\tablerefs{1. This study. 2. \citet{har07b}. 3. \citet{har07a}.}
\label{tab_CEm105}
\end{deluxetable*}

%%%%%%%%%%%%%%%%%%%%%%%%%%%%%%%%%%%%%%
%%% Fig.10         Fig 6 New 2015 Dec
%%%%%%%%%%%%%%%%%%%%%%%%%%%%%%%%%%%%%%
\begin{figure*}
\centering
\includegraphics[scale=1.0]{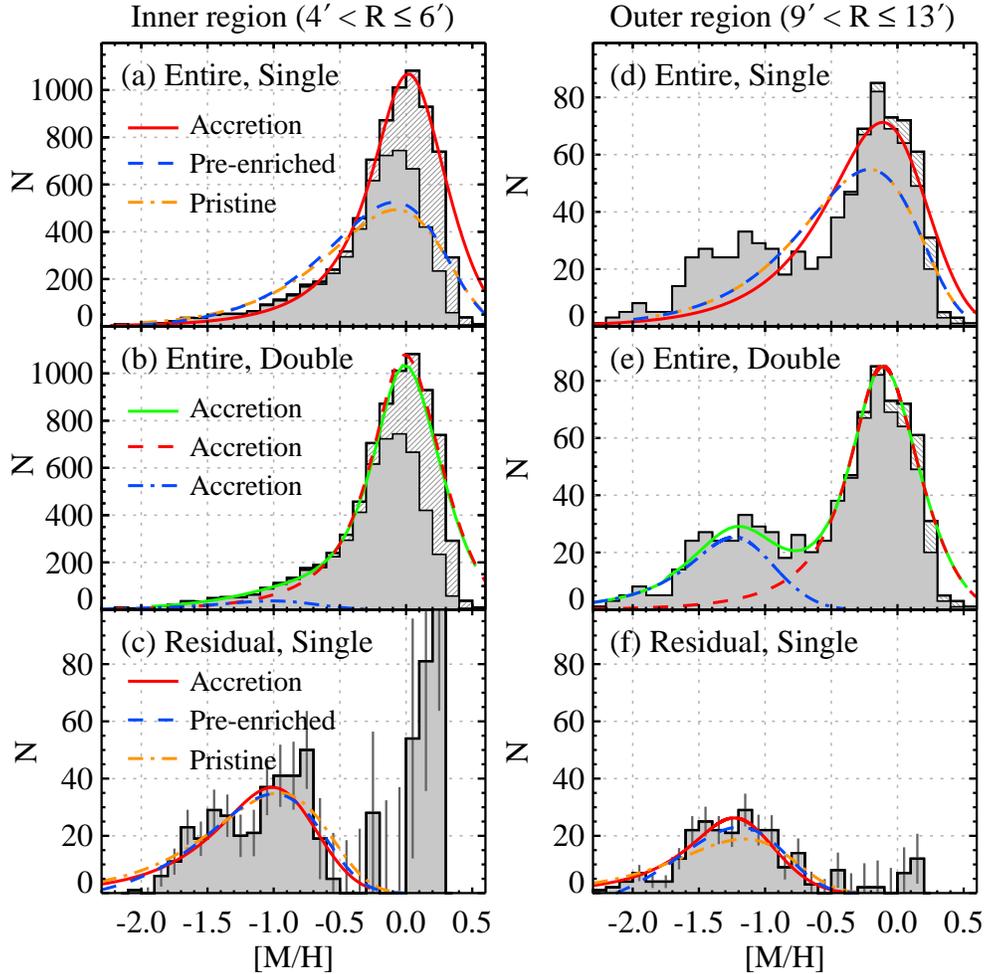} %{mdf42.eps} %{mdf0.eps} %{mdf1.eps} %{mdf_n3384.eps} 
\caption{Chemical evolution models for the MDFs of the bright RGB stars ($M_{\rm bol} \le 3.0$) in the inner region ($4\arcmin<R \le 6\arcmin$) (left panels) and outer region ($9\arcmin<R \le 13\arcmin$) (right panels) of M105. 
(Top panels) Single component models.
(Middle panels) Double component accreting gas models.
The fitting range for the chemical evolution models is [M/H]$<+0.3$.
(Bottom panels) The residual MDFs after subtraction of the metal-rich component in the accreting gas model.
Yellow dot-dashed lines, blue dashed lines, and red solid lines denote the pristine closed-box models, the pre-enriched closed-box models, and the accreting gas models, respectively. 
Fitting parameters for the models are listed in {\bf Table \ref{tab_CEm105}}.
} 
\label{fig_CEm105}
\end{figure*}

Simple analytic chemical evolution models have been used to investigate how galaxies evolve from the comparison with the MDFs of resolved stars in nearby galaxies (\citet{tal71,lyn75,tin80,pag97,har07a,har07b,kir11,pea15} and references therein).
We compared the MDFs of the resolved stars in M105 with three analytic chemical evolution models, as applied to the studies of the Local Group dwarf spheroidal galaxies by \citet{kir11}  and NGC 3115 by \citet{pea15}: a pristine closed-box model,
a pre-enriched closed-box model, and an accreting gas model.

First, the pristine closed-box model is based on the assumptions that a galaxy is a closed-box with no inflow or outflow, that gas is converted only by star formation,
and that initial gas (pristine gas) is metal-free. This model predicts an MDF:
$$N({\rm [M/H]}) \propto (10^{\rm [M/H]}) \exp ({{-10^{\rm[M/H]}} \over p} ),$$

\noindent where $p$ is a metal yield that is a ratio of the mass of metals newly produced  by the new generation of stars and the total mass in the survived stars among them. The higher the value of $p$ is, the higher the value of the peak metallicity in the MDF is.

Second, the pre-enriched closed-box model allows  initial gas to be enriched with [M/H]$_0$, to reduce the well-known G-dwarf problem. This model predicts an MDF:

$$N({\rm [M/H]}) \propto (10^{\rm [M/H]} - 10^{\rm [M/H]_0 }) \exp ({{-10^{\rm[M/H]}} \over p} ).$$

Third, the accreting gas model (the extra gas model or the best accretion model \citep{lyn75,pag97,kir11}) allows external gas to accrete to the galaxy during its evolution. In this model we can describe the relation between the fraction of the gas mass with respect to the initial gas mass $g$ 
and that of the stellar mass  $s$ by $g(s)=(1-s/M) (1 + s - s/M)$ where $M$ is the final stellar mass in units of the initial gas mass. $M$ can be larger than one in this model, while $M=1$ in the closed-box models. 
In this model the free parameters are $p$ and $M$.
 This model predicts an MDF that is much narrower than the closed-box models, as described by

$$N({\rm [M/H]}) \propto {10^{\rm [M/H]} \over p}
{{1+s(1-1/M)} \over { (1-s/M)^{-1} - 2 (1-1/M) (10^{\rm [M/H]} / p )}}. $$ 

\noindent Here the relation between $s$ and [M/H] is not simple, but is given in a complicated form by

$${\rm [M/H]}(s) = \log ( p ( {M \over {1+s-s/M}} )^2  \ln ( { 1 \over { 1-s/M}} - {s \over M} (1 - {1 \over M})  ).$$

We fit the MDFs of the bright RGB stars with $M_{\rm bol} \leq-3.0$ mag in M105 using these models, summarizing the results in {\bf Table \ref{tab_CEm105}} and {\bf Figure \ref{fig_CEm105}}. 
We tried to find the best-fit values of the parameters using least-squares minimization in the parameter space. 
For model fitting we used the MDFs for the range [M/H]$\leq0.3$ where incompleteness is not serious. 
In the case of the accreting gas models, we tried both
single component models and double component models (metal-rich one and metal-poor one), finding
that the latter fit better the MDFs than the former.
Errors for the parameters were calculated using the bootstrapping method with 1000 trials.  
%although the high metallicity parts of the MDFs are fit very well by the accreting gas models, the low metallicity parts show some excess above the model lines in the inner region, and significant excess in the outer region. 
In the double component modeling, we fit first the metal-rich part of the MDFs ($-0.6<$[M/H]$\leq0.3$) using an accreting gas model as shown by the solid lines, and subtract the resulting model from the MDF. Then the residual MDF is fit using another model. Then we subtract this model from the original MDF, and fit the it with an accreting gas model. 
We iterate this procedure until the entire MDF is fit well by the double component models. 

%  are displayed in the lower panels of the same figure. 
{\bf Figure \ref{fig_CEm105}} shows several interesting features.
First, in the case of single component models (top panels), the MDF of the inner region is fit reasonably well by the accreting gas model, but not by any of the closed-box models. 
On the other hand, the MDF of the outer region is poorly fit with any of the models. 
The MDF of the outer region shows a significant excess above the accretion model in the low metallicity range, while that of the inner region reveals a weak excess in the similar low metallicity range at [M/H]$\leq-0.7$.

Second, the MDFs of both regions are fit remarkably well by the accreting gas model with double components (middle panels). 
To show better the MDFs of the metal-poor component only, we subtracted the metal-rich component models from the original MDF, and displayed the residuals in the bottom panels.
The errorbars in the residual MDFs represent Poisson errors of the original MDFs.
It has been generally assumed in the chemical evolution models that the yield ($p$) does not vary depending on metallicity. If the yield is a function of metallicity, a single-component accreting gas model would fit the data even for the low metallicity range. However, there are no physical justifications for this and nothing is known about how it varies as a function of metallicity.

Third, interestingly the residual MDFs  show clearly an excess in the low metallicity range at [M/H]$\leq-0.5$ in both the inner and outer regions. 
The excess in the narrow high metallicity range  at [M/H]$>0.0$ for the inner region is dominated by  the errors,  as shown by the errorbars,
%considered to be dominated by the shot noise in the subtraction of two large numbers 
and also by the larger uncertainty in the metallicity estimation so that it can be ignored. 
We fit the residual MDFs for [M/H]$\leq-0.5$ with single component models, finding that all three models fit reasonably well the residual MDFs.
It is noted  also that the shapes of the residual MDFs in both regions appear to be similar to those for the bright dwarf spheroidal galaxies \citep{kir11}. 

Fourth, in the accreting gas models with double components,
the metal yield parameter for the metal-rich component ($p=(1.14\pm0.02) Z_\odot$) in the inner region is about one order of magnitude larger than that for the metal-poor component ($p=(0.10\pm0.02) Z_\odot$), 
while  the final mass parameter for the metal-rich component ($M=2.93\pm0.18$) for the inner region is about twice larger than that for the metal-poor component ($M=1.35\pm0.63$).
Similar trends are also seen in the outer region:
$p=(0.90\pm0.04) Z_\odot$ and $M=2.84\pm0.51$ for the metal-rich component, and
$p=(0.06\pm0.01) Z_\odot$ and $M=1.83\pm0.56$ for the metal-poor component.

Fifth, 
the size ratio of the metal-rich component and the metal-poor component is 21.6:1.0 for the inner region, and
2.4:1.0 for the outer region. Thus the contribution of the metal-poor component is minor (4.4\%) in the inner region, but it
becomes significant (29.4\%)  in the outer region.
%

%%%%%%%%%%%%%%%%%%%%%%%%%%%%%%%%
% Table 6                             2  6galaxiesb.tex Apr 29, 2015
%%%%%%%%%%%%%%%%%%%%%%%%%%%%%%%%
\begin{deluxetable*}{lccccccc}
\tabletypesize{\footnotesize} % \footnotesize \scriptsize
%\tabletypesize{\tiny}
\setlength{\tabcolsep}{0.05in}
%\rotate
\tablecaption{A Summary of HST Data for Nearby Early-type Galaxies}
\tablewidth{0pt}

\tablehead{ \colhead{Parameter}  & \colhead{Virgo Core Field} & \colhead{NGC 5128} &  \colhead{NGC 3377} & \colhead{NGC 404} & \colhead{NGC 5011C}  & \colhead{ESO410-005} & 
} 
\startdata
% \multicolumn{8}{c}{Description of $HST$ observation} \\
P.I.	& 10131  & 9373 & 9811 & 10915 & 12546 & 10503\\
R.A.(2000)	& 12 28 10.9 & 13 25 14.01 & 10 47 48.1& 01 09 17.0& 13 13 11.9 & 00 15 32.2 \\
Dec(2000) 	& +12 33 21 & --43 34 27.8 & +13 55 45 & +35 44 46 & --43 15 56 & --32 10 36 \\
Instrument & ACS/WFC & ACS/WFC & ACS/WFC & WFPC2 & ACS/WFC  & ACS/WFC \\
T(exp), F606W  & 63,440s  & 30,880s & 38,500s & 36,400s & 900s & 13,440s \\
T(exp), F814W  & 26,880s  & 30,880s & 22,260s & 36,400s & 900s & 26,880s \\
\hline
\enddata
%\tablerefs{1. This study. 2. \citet{har07b}. 3. \citet{har07a}.}
\label{tab_4gallog} %data}
\end{deluxetable*}

%%%%%%%%%%%%%%%%%%%%%%%%%%%%%%%%
% Table 7                            2  6galaxiesb.tex Apr 29, 2015
%%%%%%%%%%%%%%%%%%%%%%%%%%%%%%%%
\begin{deluxetable*}{lccccccccc}
\tabletypesize{\scriptsize} %\footnotesize} 
% \footnotesize \scriptsize
%\tabletypesize{\tiny}
\setlength{\tabcolsep}{0.05in}
%\rotate
\tablecaption{A Summary of Basic Information for Nearby Early-type Galaxies}
\tablewidth{0pt}

\tablehead{ \colhead{Parameter} & \colhead{M105} & \colhead{Virgo Core Field$^a$} & \colhead{NGC 5128} & \colhead{NGC 3377} & \colhead{NGC 404} & \colhead{NGC 5011C}  & \colhead{ESO410-005} & Reference& 
} 
\startdata
R.A.(2000) 	& 10 47 49.6 & 12 30 49.4 (M87) & 13 25 27.6 & 10 47 42.3 & 01 09 27.0 & 13 13 11.9  & 00 15 31.5	& 1\\
Dec(2000) 	& +12 34 54  & +12 23 28 (M87)  & --43 01 09& +13 59 09  & +35 43 05  & --43 15 56   & --32 10 48 	& 1\\
Type	& E1	& cD pec (M87) & Ep & E5-6	& SA0	& S0pec	 & dE3	& 1,2\\
$B$		& 10.24	& 9.59  & 7.78 & 11.24 & 11.21 & 14.37	 & 14.93	& 2,3\\
$B-V$	& 0.96  & 0.96  & 1.00 & 0.86  & 0.94  & 0.9$^b$ 		& 0.9$^b$& 2,3 \\
$V$		& 9.28  & 8.63  & 6.78 & 10.38 & 10.27 & 13.47$^b$	 	& 14.03$^b$ & 2,3\\
$I$		& 		& 7.92  & 	   & 		& 		& 13.25	 & 13.86	& 3\\
$E(B-V)$& 0.022	& 0.020 & 0.101 & 0.030 & 0.051 & 0.106  & 0.012 & 1,4\\
$E(V-I)$& 0.030	& 0.028 & 0.142 & 0.042 & 0.072 & 0.150  & 0.017 & 1,4\\
$A_V$	& 0.069	& 0.063 & 0.315 & 0.093 & 0.160 & 0.331  & 0.037 & 1,4\\
$A_I$	& 0.037	& 0.035 & 0.173 & 0.051 & 0.088 & 0.181  & 0.020 & 1,4\\
\hline
% \multicolumn{8}{c}{Description of $HST$ observation} \\
%$P.I.$	& 10413 (inner) & 10131  & 9811 & 10915 & 12546 & 10503\\
%$R.A.$ 	& 10 48 01.5 & 12 28 10.9& 10 47 48.1& 01 09 17.0& 13 13 11.9 & 00 15 32.2 \\
%$Dec$ 	& +12 32 18  & +12 33 21 & +13 55 45 & +35 44 46 & --43 15 56 & --32 10 36 \\
%$Instrument$ & ACS/WFC & ACS/WFC & ACS/WFC & WFPC2 & ACS/WFC  & ACS/WFC \\
%Expo, $F606W$ & 19,550s & 63,440s  & 38,500s & 36,400s & 900s & 13,440s \\
%Expo, $F814W$ & 19,550s & 26,880s  & 22,260s & 36,400s & 900s & 26,880s \\
%\hline
\multicolumn{8}{c}{Distance estimation} \\
$I_{TRGB}$ 		& $26.02\pm0.02$ & $26.97\pm0.03$ & $23.91\pm0.03$ & $26.19\pm0.02$ & $23.50\pm0.02$ & $24.00\pm0.02$  & $22.37\pm0.02$ & 5\\
$A_I$			& 0.039 & 0.038  & 0.185 & 0.050 & 0.078  & 0.181 & 0.020 & 1,4\\
$I_{0,TRGB}$ 	& $25.98\pm0.02$ & $26.93\pm0.03$ & $23.75\pm0.03$& $26.14\pm0.02$ & $23.42\pm0.02$ & $23.82\pm0.02$ & $22.35\pm0.02$ & 5\\
$E(V-I)$		& 0.032  & 0.031 & 0.152 & 0.041 & 0.063 & 0.150& 0.017 & 1,4\\
$(V-I)_{TRGB}$ 	& $1.58\pm0.03$ & $1.47\pm0.02$ & $1.88\pm0.3$ & $1.71\pm0.03$ & $1.66\pm0.02$ & $1.71\pm0.02$ & $1.43\pm0.02$ & 5\\
$M_{I,TRGB}$ 	& --4.06 & --4.09 & --4.02 & --4.04 & --4.04 & --4.06 & -4.09 & 5\\
$(m-M)_0$ 		& $30.04\pm0.02$ & $31.02\pm0.03$ & $27.75\pm0.03$ & $30.18\pm0.02$ & $27.46\pm0.02$ & $27.88\pm0.02$  & $26.44\pm0.02$ & 5\\
\hline
$M_V$ 			& --20.83 & --22.45  & --21.29 & --19.89 & --17.35 & --14.74 & --12.45 & 5 \\
\enddata
\tablenotetext{a}{The Virgo core field is located at about $40\arcmin$ ($\sim$30 $R_{\rm eff}$) from the center of M87 (see {\bf Figure 1} in \citet{jan14}). We list the parameters for M87 derived in this study.}
\tablenotetext{b}{Assumed.}
\tablerefs{(1) NED; (2) RC3 \citep{dev91}; (3) Hyperleda; (4) \citet{sch11}; (5) This study.}
\label{tab_4gal} %data}
\end{deluxetable*}

Sixth, in the accreting gas models with double components,
the peak metallicity of the metal-rich component in the inner region ([M/H]$\approx 0.0$) is only 0.1 dex higher than that for the outer region ([M/H]$\approx -0.1$). 
Interestingly the peak metallicity of the metal-poor component in the inner region ([M/H]$\approx -1.0$) is  about 0.2 dex higher than that for the outer region ([M/H]$\approx -1.2$). 
Thus the metallicity gradient of each component between the two regions (mean galactocentric distances of the stars in each component are $R=11\farcm1$ and $4\farcm5$) is weak:
 d[M/H]/dLog(R) %$ = -0.1 log(11\farcm1/4\farcm5)
 $= -0.26$ for the metal-rich component, and
 d[M/H]/dLog(R) %$ = -0.2/log(11\farcm1/4\farcm5) 
 $= -0.51$ for the metal-poor component.
In contrast, the mean value of the metallicity of the entire bright RGB stars in the MDFs is 
[M/H]  %=-0.176\pm0.418/\sqrt{7740}
 $=-0.176\pm0.005$ for the inner region 
%(at $R\approx 4\arcmin.5$), 
and [M/H] %$=-0.516\pm0.604/\sqrt{802}
 $=-0.516\pm0.021$ for the outer region % (at $R\approx 12\arcmin.0$)
(median values are $-0.050$ and $-0.350$, respectively).
%If we use $I=27$ mag for the lower limit of the sample instead of $M_{\rm bol}=-3.0$ mag, 
%the mean value %of the metallicity of the entire bright RGB stars with $I<27$ mag 
%becomes slightly lower: $[M/H]=-0.378\pm0.006$ for the inner region 
%and $[M/H]=-0.658\pm0.022$ for the outer region 
%(median values are $-0.250$ and $-0.450$, respectively).
Thus the metallicity difference between the two regions is $\Delta$[M/H]$=-0.34\pm0.02$, yielding a radial gradient of d[M/H]/dLog(R) %$= -0.34/log(11\farcm1/4\farcm5) 
$=-0.87\pm0.05$.
This value is a few times larger than the value for each of the metal-rich and metal-poor component.
Therefore the metallicity gradient of the whole RGB in M105 is %dominated by a different mixture of the metal-rich and metal-poor populations, 
mainly a population gradient, not by a gradual change of the peak metallicity of single populations.
%Metal rich    Metal poor
%15866.2    :  733.9         = 21.6:1      (Inner region)
%1209.4      :  506.2         = 2.39:1      (Outer region)
%1. Mbol < -3.0
%              Inner Region         Outer Region
%              (7740 stars)         (802 stars))
%Mean      -0.176                   -0.516              [M/H]
%Median    -0.05                    -0.350              [M/H]
%Stdev      0.418                    0.604              [M/H]

%2. I =< 27.0
%              Inner Region         Outer Region
%              (4896 stars)         (661 stars))
%Mean      -0.378                  -0.658              [M/H]
%Median    -0.250                  -0.450              [M/H]
%Stdev      0.400                    0.571              [M/H]
%}

%\clearpage

\subsection{Comparison of the MDFs of the ETGs}

\citet{har07a} presented a comparison of the MDFs for RGB stars from dwarf  galaxies (Draco, Leo I, NGC 147) to bright ETGs (NGC 3377, M105 and NGC 5128).
 They found that faint ETGs show only low metallicity stars, while
the bright ETGs show low metallicity stars with an extremely broad range of metallicity as well as high metallicity stars, and that mean metallicity is higher for brighter (more massive) ETGs. 
%\citet{har07a} Talbot, R. J., Jr., & Arnettsuggested that these metallicity distribution functions for ETGs can be explained with 
%a distinct two-stage accreting-box chemical evolution model. 
%First, metal-poor halos were formed rapidly during the initial formation phase in the closed system
%with very low effective yield ($y-{\rm eff} \approx 0.1 Z_\odot$) and truncation at low metallcity ($Z \approx 0.2 Z_\odot$).
%Second, metal-rich components  were formed during the accreting phase in the massive systems with high effective yield
%($y-{\rm eff} \approx 0.5 Z_\odot$). 
\citet{mou10} found  that the MDF of the RGB stars in the halo field of a very bright Sa galaxy, M104 (NGC 4594, Sombrero), is similar to those for the ETGs with similar luminosity, 
indicating that the chemical evolution of the halo in the disk galaxies may be similar to that of the ETGs.  

%the metal-poor, diffuse halo be- longed to a rapid initial formation phase strongly resembling a
%simple closed-box model with very low effective yield 
%yeff ~ 0.1 Z and truncated at abundance Z~0.2Zo 
%By contrast, the metal-rich component (the centrally concentrated spheroid or,
%possibly, the nearly face-on disk if NGC 3379 is actually an S0)
%requires an accreting-box type of model and an effective yield
%5 times higher.

We selected four ETGs with a wide range of luminosity to compare with M105:
NGC 3377 (E5, $M_V=-19.89$), % \citet{har07b}),
 NGC 404 (dS0, $M_V=-17.35$),  %\citet{har09b}),
 NGC 5011C (dS0, $M_V=-14.74$),  %\citet{sav07}) , 
and
ESO410-005 (dE3, $M_V=-12.45$). %, \citet{tik13}).
Resolved stars in these galaxies were studied previously:
NGC 3377 \citep{har07b},
NGC 404 \citep{wil10},
NGC 5011C \citep{sav07}, and
ESO410-005 \citep{dac10,tik13,yan14}.  
We also included one halo field at $R\approx 33\arcmin$ of NGC 5128 previously studied by \citet{rej05}. NGC 5128  (Ep, $M_V=-21.29$) is 0.5 mag  brighter than M105.

In addition, we selected a field close to the Virgo center where there are no bright galaxies (called the Virgo core field). This  field is located at about $40\arcmin$ ($\sim$30 $R_{\rm eff}$) from the center of M87 (see {\bf Figure 1} in \citet{jan14}). 
This field is far from nearby bright galaxies (including M87, M86 and M84), representing an intracluster field close to the Virgo center. 
\citet{wil07} presented a study of the intracluster stars in this field based on the HST F606W and F814W imaging. 
They  suggested in the study of resolved stars in the Virgo core field that the population younger than 10 Gyr is more metal-rich than that older than 10 Gyr. They used Starfish CMD model fitting \citep{har01} and the isochrones given by \citet{gir02} to derive the star formation history in this field. They found that this field is dominated by low metallicity old stars, but includes a small fraction of younger popoulation with high metallicity. Thus the MDF of this field we derived with 12 Gyr isochrones may be affected by the presence of younger population in this field, but the effect may be minor.

A summary of HST data and basic properties of these galaxies are listed in {\bf Table \ref{tab_4gallog}} and {\bf \ref{tab_4gal}}.
%Basic properties of these galaxies are listed in {\bf Table \ref{tab_4gal}}.
We obtained photometry of the resolved stars in each galaxy from the F606W and F814W images in the HST archive, using the same procedures as done for M105. 
We determined the distances to these galaxies using the TRGB method.

%%%%%%%%%%%%%%%%%%%%%%%%%%%%%%%%%%%%%
%%  Fig11
%%%%%%%%%%%%%%%%%%%%%%%%%%%%%%

\begin{figure}
\centering
\includegraphics[scale=0.8]{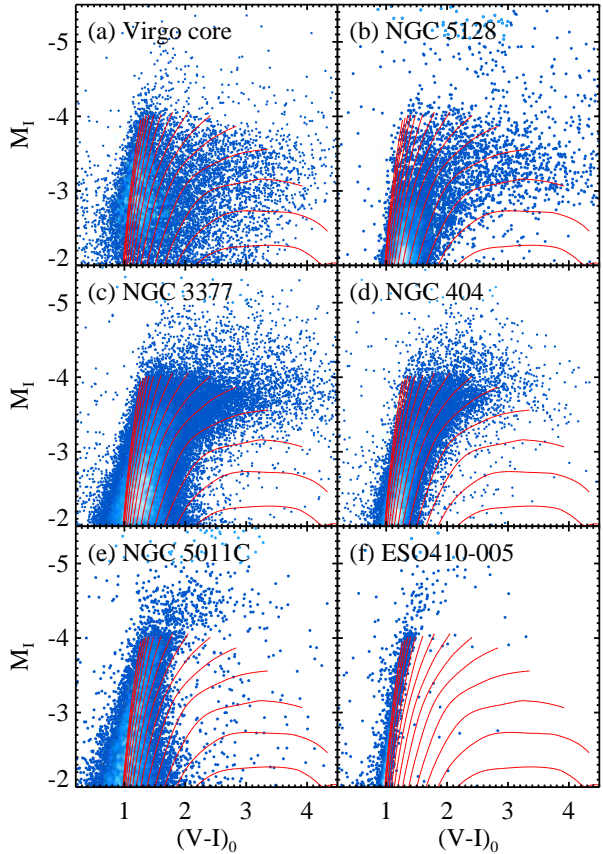} %{fig9.eps}%{cmd5.eps} %{cmd42.eps}
\caption{$M_I - (V-I)_0$ CMDs of the resolved stars in the Virgo Core Field, 
NGC 5218 (Ep, $M_V=-21.29$), NGC 3377 (E5, $M_V=-19.89$), NGC 404 (S0, $M_V=-17.35$), NGC 5011C (S0, $M_V=-14.74$), and ESO410-005 (dE3, $M_V=-12.45$).
Red solid lines represent 12 Gyr isochrones for [$\alpha$/Fe]=+0.2 and a range of metallicity ([M/H] = --2.2 to +0.4) with a step of 0.2 from left to right in the Dartmouth models \citep{dot08}.} 
\label{fig_cmdcomp} %42}
\end{figure}

\begin{figure}
\centering
\includegraphics[scale=0.9]{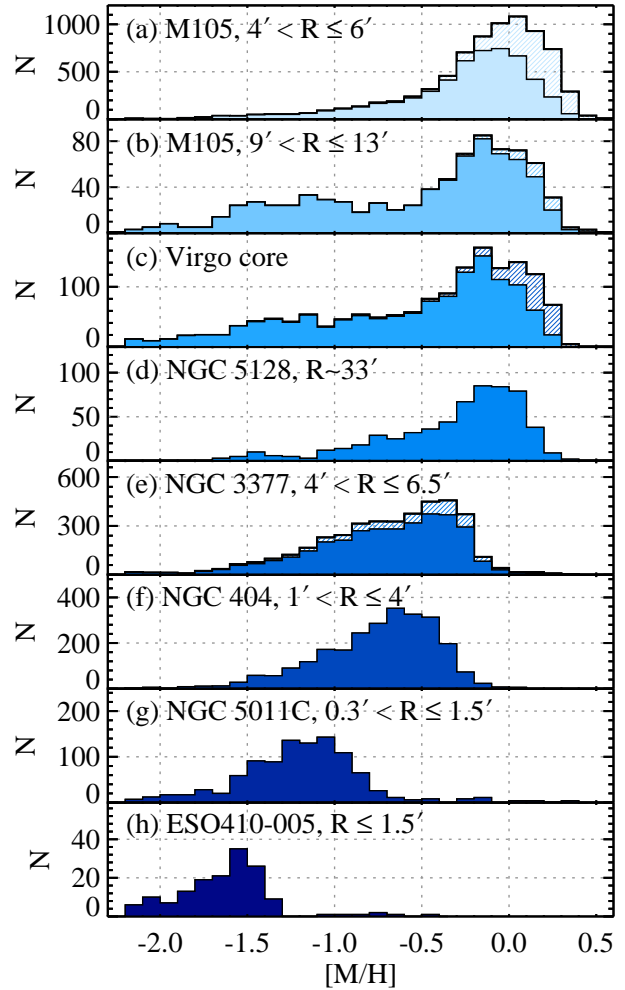} %{mdf7.eps} %{mdf62.eps}
\caption{A comparison of MDFs of the resolved RGB stars in M105, the Virgo Core Field, 
NGC 5128 (Ep, $M_V=-21.29$), 
NGC 3377 (E5, $M_V=-19.89$), 
NGC 404 (S0, $M_V=-17.35$), 
NGC 5011C (S0, $M_V=-14.74$), and ESO410-005 (dE3, $M_V=-12.45$).
The filled and hatched histograms %for M105 and NGC 3377 
represent the MDFs before and after completeness correction, respectively.
Completeness corrections for other MDFs are negligible, as shown by their CMDs.} 
\label{fig_mdfcomp} %62}
\end{figure}

%%%%%%%%%%%%%%%%%%%%%
% FIg 11
%%%%%%%%%%%%%%%%%%%%%
{\bf Figure \ref{fig_cmdcomp}} displays the CMDs of the resolved stars in these galaxies and the Virgo core field . Absolute magnitudes and intrinsic colors of the resolved stars were derived, adopting the distances derived in this study and foreground reddening values \citep{sch11} for each galaxy.
We also overlayed 12 Gyr isochrones for [$\alpha$/Fe] = 0.2 and a range of metallicity ([M/H] = --2.2 to 0.4) \citep{dot08}.
Then we derived the metallicity of the resolved stars using the same procedures as done for M105.
In {\bf Figure \ref{fig_cmdcomp}} all galaxies show prominent RGBs. However, the width of the RGB gets narrower and the mean color of the RGB gets bluer, as their host  galaxies become fainter.
%We derived the MDFs of these stars in the galaxies
%using the same procedures as done for M105.

%%%%%%%%%%%%%%%%%%%%%%%%%%%%%%%%%%%%%
%%  Fig 11
%%%%%%%%%%%%%%%%%%%%%%%%%%%%%%%%%%%%%
\begin{figure*}
\centering
\includegraphics[scale=1.0]{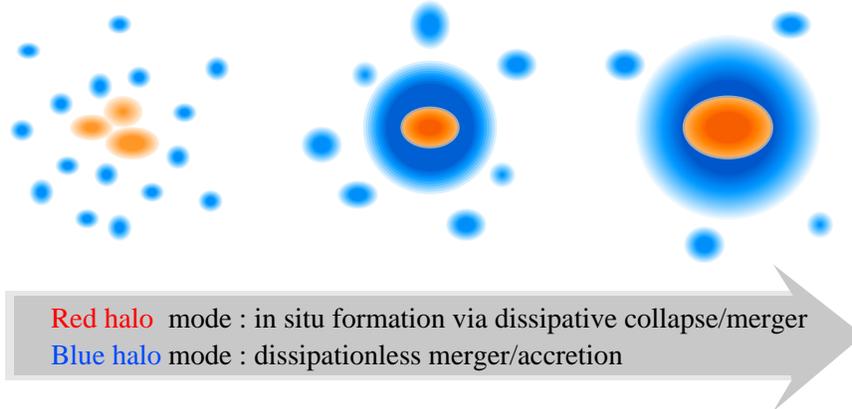} %{fig11.eps} %{scenario.eps}
\caption{ A schematic sketch for the dual halo mode formation scenario to explain how massive ETGs are formed.
In the first phase of the red halo mode, red halos dominated by metal-rich stars are formed in the massive progenitors (either a single massive protogalactic cloud or merger of two or  more massive galaxies), while blue halos of metal-poor stars are formed in the low-mass galaxies around the massive progenitors more than 10 Gyrs ago. 
In the second phase of the red halo mode, red halos grow via merger of intermediate-to-high mass galaxies, becoming larger and more massive. 
In the blue halo mode, the blue halos around the massive red halos grow via dry mergers of low-mass galaxies for long until today. 
} 
\label{fig_scenario}
\end{figure*}

In {\bf Figure \ref{fig_mdfcomp}} we compare the MDFs for the bright RGB stars %($I\leq 27$) 
in the inner and outer regions of M105 
($4\arcmin<R\leq6\arcmin$ and $9\arcmin<R\leq13\arcmin$) and those in other galaxies and the Virgo core field.
%We also derived  the MDF for the bright RGB stars  in NGC 3377 using the same procedure, from the photometry we obtained from the HST/ACS images, as displayed in  {\bf Figure \ref{fig_cmdn3377}}. 
%displays the CMDs of the RGB stars at $1\farcm7\lesssim R\lesssim5\farcm9$ regions in NGC 3377. 
%Our photometry of NGC 3377 stars goes about one magnitude deeper than that given in \citet{har07a} (see their Fig. 6), covering redder (more metal-rich) part of the RGB more.
%The effective radius of NGC 3377 is $R_{\rm eff, NGC 3377} = 0\arcmin.93 = 2.8 $ kpc \citep{dev91}, similar to that of M105,  so that the covered radial range corresponds to $1.8 R_{\rm eff} \lesssim R \lesssim 6.3 R_{\rm eff} $. 
%{\color{red}
%We derived the TRGB magnitude and distance for NGC 3377,$I_{TRGB}=26.19\pm0.02$ and $(m-M)_0=30.18\pm0.2\pm0.12$,}
%which is consistent with those based on the same data by \citet{har07a}. 
%We plotted the resulting MDF of NGC 3377 stars at $4\arcmin<R\leq6\farcm5$ {\color{red}} in {\bf Figure \ref{fig_mdfall}}.
%{\bf Figure \ref{fig_mdfcomp}}  shows 
 Several distinguishable features are found in this figure. 
%{\color{red} TO REMOVE! First, 
%the peak positions of the metal-rich component in the inner  and outer regions of M105 are similar ([M/H] $\approx -0.1$ and 0.0, respectively). 
%Note that the ratio of the total numbers of the metal-poor and metal-rich stars in Table 4 increases from 0.4 to 0.7 as the galactocentric distance increases. 
%This results in the radial gradient of the metallicity.
%Note that the radial gradient of the mean metallicity is  mainly due to a different population mixture rather than due to a gradual change of the peak metallicity.}
%%
First, %Second, 
the MDF of the outer region in M105 is remarkably similar to that of the intracluster field in the Virgo. Both MDFs show two major components and their peak positions are similar ([M/H] $\approx -1.2$ and --0.1, respectively). 
This shows that the outer field of massive galaxies and intracluster field close to the center of galaxy clusters may share a similar history of growing. 
The existence of the significant metal-rich component with [M/H] $\approx -0.1$ in both the outer halo of M105 and the Virgo core field implies that a significant fraction of the progenitors for the stars in these low density regions are massive galaxies (as massive as M105).
Second, %Third,
the peak metallicity of the high metallicity component in each galaxy is higher for brighter galaxies, consistent with the results in \citet{har07a}.
Third, %Fourth,
the peak metallicity of the metal-poor component in M105
(as well as in the Virgo core field)  is similar to that of
NGC 5011C ($M_V = -14.7$ mag). This implies that the progenitors of these metal-poor components are probably dwarf galaxies with $M_V = -14$ to --15 mag.
Fourth, %Fifth, 
%the progenitors of the metal-poor stars at the very low metallicity tail ([M/H]$<-1.5$) in M105 are probably fainter dwarf galaxies with $M_V >-12.0$ mag (dSphs and UFDs).
the progenitors of the metal-poor stars at the very low metallicity tail ([M/H]$<-1.5$) in M105 are, in part, fainter dwarf galaxies with $M_V >-12.0$ mag (dSphs and UFDs). Some of them may have come from the early stages of star formation in the central galaxy.

\subsection{Dual Stellar Halos and Formation of Massive ETGs}

Key results found in this study are
1) that the MDF of the bright RGB stars in the large range of galactocentric distance of M105 is  bimodal, 
2) that the metal-poor component is getting more dominant in the outer region of a galaxy,
3) that the radial number density profiles of the metal-rich and metal-poor stars in M105 are significantly different in the outer region,
4) that the radial number density profiles of the metal-rich and metal-poor stars in M105
%( and NGC 3377, a highly elongated elliptical galaxy (E5), 
are fitted well by the Sersic law with $n\approx 2.8$ and $n\approx 6.9$, respectively,
5) that the peak metallicity of the metal-rich component in the outer halo of M105
  is similar to that of  the Virgo core field,
6) that the peak metallicity of the metal-poor component in M105
(as well as in the Virgo core field)  is similar to that of
NGC 5011C ($M_V = -14.7$ mag), and 
%This implies that the progenitors of these metal-poor components are probably dwarf galaxies with $M_V = -14$ to --15 mag.
%
7) that the progenitors of the metal-poor stars at the low metallicity tail ([M/H] $<-1.5$) in M105 are %probably 
possibly fainter dwarf galaxies with $M_V >-12.0$ mag (dSphs and UFDs).
%3) in the highly elongated elliptical galaxy (E5), NGC 3377, the spatial distribution of the meta-poor RGB stars shows an almost circular distribution,  while that of the metal-rich RGBs stars shows a highly elongated distribution, similar to that of the stellar light
%of NGC 3377.

%The MDFs and radial distributions of the RGB stars in M105 in this study show 
These results indicate that there are two distinct stellar halos in massive ETGs like M105: a metal-poor (blue) halo and a metal-rich (red) halo. The metal-rich halo is dominant in the inner region, while the metal-poor halo is more distinguishable in the outer region.
 \citet{har07b} found the existence of a distinguishable blue RGB  in the study of the same outer field as used in this study,  %at 12 $R_{e}$ from the center of  M105 (M105-W Field), 
 and predicted that "large E/S0 galaxies
in general will have diffuse, very low-metallicity halo components, but that photometry at radii ($R \sim  (10-15) R_{\rm eff} $) will be necessary to find them''. 
Since then these diffuse low-metallicity halo components have been found in the outer regions of a few E/S0 galaxies.

%The results for the outer region in this study is consistent with those in \citet{har07b}.

The existence of dual stellar halos in massive ETGs %M105 and NGC 3377 
is consistent
with the prediction based on the globular cluster systems
in ETGs \citep{par13}: massive ETGs may have often dual halos, one blue (metal-poor) halo and one red (metal-rich)  halo. 
Considering the properties of red and blue globular cluster systems in massive ETGs, \citet{par13} suggested a dual halo scenario for how massive ETGs formed.
 Massive ETGs are formed in two main modes: a metal-rich halo mode and a metal-poor halo mode.

We provide an updated scenario, including the information we found from the stellar halos in this study. 
{\bf Figure \ref{fig_scenario}} %Figure \ref{fig_scenario}
illustrates a schematic view of the main features of this scenario. 
This scenario is based on the concept of hierarchical merging models for structure formation (\citet{cot98, del07,con14,coo15,hir15,som15} and references therein), focusing on the observational results on the properties of the resolved stars and globular clusters in ETGs.
All galaxies should have different formation histories, because they start with progenitors with different masses and are involved with diverse histories of merging. 
However, considering the bimodality of the MDFs of the resolved stars in ETGs, we can divide the formation modes roughly into two: a metal-rich halo mode and a metal-poor mode.

In the metal-rich halo mode, metal-rich halos formed in two phases. 
In the first phase, during the early phase of the universe, progenitors of today's massive galaxies formed via rapid collapse of massive protogalactic clouds (in situ formation of stars) \citep{egg62} and/or major mergers, while a much larger number of low mass galaxies and globular clusters formed via collapse of low mass clouds embedded in low mass dark matter halos mostly around the massive galaxies \citep{sea78, whi78}. These two processes (in situ collapse and major mergers) in the massive progenitors are not independent, but both can happen in the same galaxy with a small time lag. Major mergers can be either wet or dry and they can be involved also with several mergers.
In this mode, the MDF of the stars in the massive galaxies will show a dominant metal-rich component with a weak tail in the low metallicity end, as described in classical chemical evolution models. The larger the masses of the most massive progenitors are, the higher the peak values of the stellar metallicity are. The structure, color, and luminosity of the resulting galaxies are dominated by the metal-rich stars. 
In the second phase, the metal-rich halos grow mostly via merging with intermediate mass galaxies that have metallicities higher than those of dwarf galaxies. % and lower than those of massive galaxies. 
The MDFs in the resulting galaxies will be broader toward the low metallicity. They will be involved with numerous merging with low-mass/dwarf galaxies as well, but it will contribute only a minor role in increasing the mass and size of the metal-rich halos.  Thus the metal-rich halos  formed and evolve in two phases, being dominated by metal-rich stars.

In the metal-poor halo mode, metal-poor halos grow mainly via accretion or minor mergers of dwarf galaxies for long duration. The outstanding existence of a distinct blue (metal-poor) RGB in the outer region of massive galaxies shows that they are not the low metallicity tail of the main population (the metal-rich halo), but they have a distinct origin. The origin of the metal-poor stars is considered to be accreted dwarf galaxies that have mainly low metallicity stars with [M/H]$<-1.0$. Some brighter dwarf galaxies like NGC 5011C ($M_V \sim -15.0$ mag) produce the metal-poor component peak at  [M/H]$\sim -1.2$.
Among the dwarf galaxies, ultra-faint dwarf galaxies (UFDs) are one of the strong candidates for the low metallicity tail at [M/H]$<-1.5$.
UFDs have been found mostly in the Local Group \citep{bel13}. 
%Recently \citet{jan14} found the first  UFD beyond the Local Group, which is located in the intracluster region of the Virgo core. 
The stellar populations in several nearby UFDs in the Local Group \citep{bro14} and the Virgo cluster \citep{jan14} are known to be old and metal-poor, consistent with the halo stars in the low metallicity tail in M105. 
These dwarfs are accreted in the outer region of a galaxy, contributing to the growth of the metal-poor halo. Thus the accreted stars in the metal-poor halo must be very old, although they join the metal-poor halo during the long period.  
The number density of the metal-poor stars is very low so that
they make only a minor contribution to the luminosity of the galaxies.
However, they contain critical clues to understand how and when the outer halo formed and to study the distribution of the dark matter.

This dual halo mode scenario, mainly based on observational results of resolved red giants and globular clusters in nearby galaxies, is consistent with the two-phase scenario derived from high-resolution cosmological simulations  (\citet{ose10, ose12,hir15} and references therein). 
In the first phase of this scenario, compact cores of massive galaxies form rapidly via in-situ star formation associated with cold flows and/or gaseous mergers early at $2<z<6$.
These compact massive galaxies will have, on average, 
$M\approx 10^{11} M_\odot$, $R_{\rm eff} \approx 1$ kpc, and $\sigma_v  \approx 240$ \kms , around the end of this phase, and they will show disk-like structures flatter than today's ETGs.
%($M\approx 10^11 M_\odot$, $R_{\rm eff} \approx 1$ kpc, $sigma_v \approx 240$ \km)
Then in the second phase, these compact massive galaxies get bigger mainly via minor dry merging. The effective radii of these galaxies grow significantly faster in size than in mass as a result of conservation of angular momentum, and they will have lower velocity dispersion in the central region. As time goes on, these galaxies will have larger outer envelopes, which  also grow in mass, occupying a major fraction of mass in more massive galaxies. However, note that there are other views in the literature.
% (e.g., see \citet{gra15}).
For example, noting the abundance of compact massive bulges in nearby disk galaxies, \citet{gra15} concluded that not all compact massive galaxies at high redshift grow dramatically in size.  Instead some of them keep their compactness, being embedded in disks that grow slowly for long.
This scenario has been successful in explaining several observational results on the integrated properties of galaxies \citep{ose12}. 
The results on the resolved stars in M105 and other ETGs in this study provide another strong constraint to test the formation scenarios based on cosmological simulations. In particular the bimodal MDFs of the resolved stars in ETGs including these galaxies need to be considered in the future simulations.

\section{Summary and Conclusion}

Using deep photometry of the resolved stars in the large range of galactocentric distance (
%$3\arcmin \lesssim R \lesssim13\arcmin$,
$3 R_{\rm eff}  \lesssim R  \lesssim 13 R_{\rm eff}$) of M105, a standard elliptical galaxy, we provided strong evidence  that there are two subpopulations of RGB stars in this galaxy and investigated their structure and MDFs. We also presented the MDFs of the bright RGB stars in other nearby ETGs with a range of luminosity as well as in the Virgo core field derived using the same procedures as used for M105. Then we described the MDFs of M105 using the chemical evolution models, and discussed the implications of the results in terms of the dual halo mode formation scenario to explain how massive ETGs formed. 
Primary results are summarized as follows.

\begin{itemize}

\item The resolved stars in M105 are mostly RGB stars.
We measured the TRGB magnitudes  using an edge-detection method for the inner and outer region, obtaining $I_{\rm TRGB}=26.02\pm0.02$. From the mean of the inner and outer regions, we obtain a distance modulus 
$(m-M)_0=30.05\pm0.02 ({\rm random})\pm0.12 ({\rm systematic})$. This confirms that M105 is a member of the Leo I Group.

\item The bright RGB stars in the CMDs and their color distributions show the existence of two distinct subpopulations: a strong red RGB with a broad color range and a weaker blue RGB with a narrower color range. 

\item We derived metallicities of the bright RGB stars from the comparison of their photometry with theoretical isochrones for 12 Gyr age.
The MDF of the RGB stars  in M105 also shows the existence of two distinct subpopulations: a dominant metal-rich population (with a peak at [M/H]$\sim0.0$) and a much weaker metal-poor population  (with a peak at [M/H]$\sim-1.1$).

\item We divided the bright RGB stars into two groups according to their metallicity: a metal-poor one ($-2.4<$[M/H]$\leq-0.7$) and a metal-rich one ($-0.7<$[M/H]$\leq+0.5$). The radial number density profile of the metal-rich RGB stars is significantly steeper than that of the metal-poor RGB stars in the outer region, while the slopes of both are similar in the inner region.
The number density of the metal-rich stars is much higher than that of the metal-poor stars in the inner region at $R<12\arcmin$, but becomes lower thereafter.
The  radial number density profiles of the metal-rich and metal-poor RGB stars in the entire range of galactocentric distance are fit well by a S\'ersic law 
with  $n=2.75\pm0.10$  and  $n=6.89\pm0.94$, 
%$n=1.5\pm0.2$  and $n=2.0\pm0.5$, 
%with $n=1.9\pm0.3$  and  $n=2.8\pm0.8$
respectively. 
%The latter is also fit well by a single power law,
%($\sigma \propto R^{-2.5}$) with an index of $N=2.5$.
They are also fit by a single power law:
$\sigma \propto R^{-3.83\pm0.03}$ for the metal-rich stars and $\sigma \propto R^{-2.58\pm0.03}$ for the metal-poor stars.
\item
%{%\color{red}\bf 
The MDFs of both the inner region and the outer region can be described remarkably well by accreting gas models of chemical evolution with two components, a metal-rich one and a metal-poor one. The peak metallicity of the metal--rich component in the inner region is only 0.1 dex higher than that for the outer region.
The metal-poor components also show a similar trend.
The radial gradient of the metallicity between the two regions is dominated by the population mixture, rather than by the continuous change of the peak metallicity. %}

\item 
These results provide strong evidence that there are two types of stellar halos in M105: a metal-poor (blue) and a metal-rich (red) halo. 
This is consistent with the prediction in the dual halo scenario based on the spatial distribution of the globular clusters in bright ETGs by \citet{par13}.

\item Comparing the MDFs of the resolved stars in M105 with those of the resolved stars in faint to bright ETGs and in the intracluster field of the Virgo core, we suggested a dual halo mode scenario for the formation of massive ETGs.

\end{itemize}

\bigskip
The authors are thankful to anonymous referee for her/his useful comments which improved significantly our original manuscript.
This work was supported by the National Research Foundation of Korea (NRF) grant funded by the Korea Government (MSIP) (No. 2012R1A4A1028713). %BRL (No. 2013...). % Core
This paper is based on image data obtained from the Multimission Archive at the Space Telescope Science Institute (MAST).

\end{document}